\documentclass[floats,floatfix,showpacs,amssymb,prd,twocolumn,superscriptaddress,nofootinbib,nolongbibliography,reprint]{revtex4-2}

\usepackage{amssymb,amsmath,mathtools,enumitem,etoolbox,graphicx, physics,microtype,afterpage,gensymb,tabularx,xspace, bm,multirow,soul,braket}
\usepackage[dvipsnames, usenames]{xcolor}
\definecolor{linkcolor}{rgb}{0.0,0.3,0.5}
\usepackage[unicode, colorlinks=true, linkcolor=linkcolor, citecolor=linkcolor, filecolor=linkcolor,urlcolor=linkcolor, pdfusetitle]{hyperref}
\usepackage[all]{hypcap}
\usepackage[T1]{fontenc}
\usepackage[utf8]{inputenc}
\usepackage{tensor}
\usepackage{orcidlink}

\newcommand{\D}[1]{\ensuremath{\mathrm{d}}}
\newcommand{\temop}{t_{\rm EMOP}}
\newcommand{\chis}{\chi_{\rm s}}
\newcommand{\chia}{\chi_{\rm a}}
\newcommand{\phik}{\phi_{\rm k}}
\newcommand{\thetaf}{\theta_{\rm f}}
\newcommand{\chif}{\boldsymbol{\chi}_{\rm f}}
\newcommand{\epsNR}{\varepsilon^{\rm \scriptscriptstyle NR}_{\scriptscriptstyle lm}}
\newcommand{\epsGPR}{\varepsilon^{\rm \scriptscriptstyle GPR}_{\scriptscriptstyle lm}}
\newcommand{\epsGPRtot}{\varepsilon^{\rm \scriptscriptstyle GPR}_{\rm \scriptscriptstyle tot}}
\newcommand{\epsNRtot}{\varepsilon^{\rm \scriptscriptstyle NR}_{\rm \scriptscriptstyle tot}}
\newcommand{\mismGPRtot}{\mathcal{M}^{\rm \scriptscriptstyle GPR}_{\rm \scriptscriptstyle tot}}
\newcommand{\mismNRtot}{\mathcal{M}^{\rm \scriptscriptstyle NR}_{\rm \scriptscriptstyle tot}}

\newcommand{\hlc}[2][yellow]{{\colorlet{foo}{#1}\sethlcolor{foo}\hl{#2}}}

\newcommand{\bham}{{School of Physics and Astronomy \& Institute for Gravitational Wave Astronomy,\\University of Birmingham, Birmingham, B15 2TT, United Kingdom}}
\newcommand{\milan}{{Dipartimento di Fisica ``G. Occhialini'', Universit\'a degli Studi di Milano-Bicocca, Piazza della Scienza 3, 20126 Milano, Italy}}
\newcommand{\infn}{{INFN, Sezione di Milano-Bicocca, Piazza della Scienza 3, 20126 Milano, Italy}}
\newcommand{\como}{{Como Lake Centre for Astrophysics, DiSAT, Universit\'a degli Studi dell’Insubria, via Valleggio 11, 22100 Como, Italy}}

\begin{document}

\title{Ringdown mode amplitudes of precessing binary black holes}

\author{Francesco Nobili$\,$\orcidlink{0009-0008-8769-4557}}
\email{fnobili@uninsubria.it}
\affiliation{\como} \affiliation{\milan}
\affiliation{\infn} 

\author{Swetha Bhagwat$\,$\orcidlink{0000-0003-4700-5274}}
\email{s.bhagwat@bham.ac.uk}
\affiliation{\bham}

\author{Costantino Pacilio$\,$\orcidlink{0000-0002-8140-4992}}
\affiliation{\milan} \affiliation{\infn} 
\date{\today}

\author{Davide Gerosa$\,$\orcidlink{0000-0002-0933-3579}}
\affiliation{\milan} \affiliation{\infn} 
\date{\today}

\begin{abstract}
The ringdown phase of a binary black-hole merger encodes key information about the remnant properties and provides a direct probe of the strong-field regime of General Relativity. While quasi-normal mode frequencies and damping times are well understood within black-hole perturbation theory, their excitation amplitudes remain challenging to model, as they depend on the merger phase. The complexity increases for precessing black-hole binaries, where multiple emission modes can contribute comparably to the ringdown. In this paper, we investigate the phenomenology of precessing binary black hole ringdowns using the SXS numerical relativity simulations catalog. Precession significantly impacts the ringdown excitation amplitudes and the related mode hierarchy.
Using Gaussian process regression, we construct the first fits for the ringdown amplitudes of the most relevant modes in precessing systems.
\end{abstract}

\maketitle

\section{Introduction}
\label{sec:intro}

Ringdown is the final phase of a black hole (BH) merger, during which the newly formed, distorted remnant emits gravitational waves (GWs) as it settles into a stable Kerr BH~\cite{Kerr:1963ud,Teukolsky:2014vca}. Within the context of BH perturbation theory, ringdown can be modeled as a linear superposition of a countably infinite set of damped sinusoids, known as the quasi-normal mode (QNM) spectrum~\cite{Teukolsky:1972my,Teukolsky:1973ha}. According to the uniqueness theorems~\cite{Israel:1967wq,Hawking:1971vc,Carter:1971zc,Robinson:1975bv}, the frequencies and damping times of QNMs are fully determined by the mass and spin of the BH~\cite{Vishveshwara:1970zz,Press:1971wr,Press:1973zz}. Testing this prediction forms the foundation of the the ``BH spectroscopy'' research program~\cite{Dreyer:2003bv,Berti:2005ys,Berti:2007zu,Meidam:2014jpa,Berti:2018vdi}. 

The amplitudes and phases of the modes carry information that is complementary to that contained in the frequencies and damping times. As in any perturbation theory, the (quasi-)normal mode frequencies are intrinsic properties of the perturbed system, whereas the specific modes that are excited, along with their amplitudes, depend on the nature of the initial perturbation. For a BH ringdown, excitation amplitudes are predominantly determined by the dynamics during the plunge and merger phases~\cite{Bhagwat:2017tkm}. This is the most complex stage of binary BH evolution, dominated by strong-field, nonlinear dynamics, making it a promising target for probing potential deviations from General Relativity (GR). While tests of GR based on mode amplitudes have historically received less attention than spectroscopy-based approaches, new frameworks have recently been proposed and are now under active development~\cite{Forteza:2022tgq, Capano:2021etf}. 
Crucially, testing GR using excitation amplitudes requires accurate predictions across the parameter space of binary BH signals.

Ringdown amplitude predictions have broader applications in gravitational waveform modeling. For instance, for several state-of-the-art approximants (e.g., the Phenom and EOB families), the ringdown phase needs to be seamlessly attached to the earlier inspiral (modeled semi-analytically) and merger (calibrated to numerical simulations) phases, ensuring that the resulting signal is sufficiently smooth.

Several numerical-relativity (NR) fits are now available for predicting mode amplitudes in a specific subset of binary BH systems, namely those where the pre-merger BH spins are aligned with the orbital angular momentum of the binary~\cite{Kamaretsos:2012bs,London:2014cma,Baibhav:2017jhs,London:2018gaq,Borhanian:2019kxt,Forteza:2022tgq,Cheung:2023vki,MaganaZertuche:2024ajz,Pacilio:2024tdl, Carullo:2024smg}.
Notably, in Ref.~\cite{Pacilio:2024tdl}, we recently presented a Gaussian Process Regression (GPR) model for predicting QNM amplitudes and phases in the spin-aligned scenario. In this paper, we present a comprehensive investigation of the phenomenology of ringdown amplitudes for BH binaries with misaligned spins and introduce the first %
 NR-calibrated approximant capable of capturing such systems.

When the BH spins are misaligned, both the orbital angular momentum and the spins precess, altering their orientation as the binary evolves. This significantly modifies the perturbation conditions for the final remnant BH, leading to substantial changes in the ringdown amplitudes~\cite{Finch:2021iip,Zhu:2023fnf}.
 As shown in this paper, the variation induced by precessing spins can be as large as $100\%$!
Considering an expansion in spherical harmonics, we provide fits for the modes $(l,m,n) = (2, \pm2,0), (2, \pm1,0), (2, 0,0)$, and $(3, \pm 3,0)$.
Specifically, we provide two sets of GPR regression fits. First, we consider a physically motivated 6-dimensional parameter space, which %
 considers suitable combinations of the BH masses and spins, along with the directions of the remnant spin and proper velocity (or ``kick'')~\cite{Zhu:2023fnf}. Additionally, since most parameter estimation and waveform modeling pipelines are currently set up to take inputs in terms of mass ratio and BH spins, we also provide regression results for the 7-dimensional parameter space consisting of the mass ratio and the two BH spin vectors.  Our models are released as part of the \textsc{postmerger} open-source package for the python programming language~\cite{Pacilio_postmerger_code}.

Our paper is structured as follows.
In Sec.~\ref{sec:training_dataset}, we introduce the NR dataset used and the necessary pre-processing steps. In Sec.~\ref{sec:amp_phenom}, we explore the phenomenology of precessing ringdown amplitudes. In Sec.~\ref{sec:mode_amp_fits}, we illustrate and benchmark our GPR fits. Concluding remarks are presented in Sec.~\ref{sec:conclusions}.
Hereafter, we indicate the BH masses with $m_{1,2}$ and their dimensionless spin parameters vectors $\boldsymbol{\chi}_{1,2}$.

\section{Mode amplitude extraction}
\label{sec:training_dataset}

\subsection{Numerical-relativity catalog}
\label{sec:conventions}

We extract ringdown amplitudes from NR simulations, specifically using the simulations SXS:BBH:0001-2265 from the version of the SXS public catalog provided in Ref.~\cite{Boyle:2019kee}. We select simulations as follows:

\begin{itemize}

\item We restrict our analysis to quasi-circular mergers, excluding simulations with eccentricity $e > 10^{-3}$. In the SXS catalog, the eccentricity after relaxation time is provided using a linear-order Newtonian approximation; in some cases, only an upper limit is available, which we conservatively adopt as the eccentricity value. We observe that mild to moderate eccentric simulations do not exhibit significant deviations in QNM amplitude trends compared to quasi-circular simulations with similar parameters~\cite{Forteza:2022tgq}. Reference~\cite{Carullo:2023kvj} shows that noticeable deviations from the quasi-circular case occur only for eccentricities $e_0 \gtrsim 0.5$, and our dataset does not include such high eccentricities. We have explicitly verified that including these eccentric simulations in the GPR training set does not notably impact model performance.

\item We exclude seven other simulations due to technical issues, such as differences in their conventions or the presence of numerical artifacts. Specifically, simulations SXS:BBH:0171, SXS:BBH:1134, and SXS:BBH:0170 have remnant masses larger than the total initial mass; SXS:BBH:1131 has inconsistencies in the reported mass ratio; SXS:BBH:1110 exhibits strong numerical artifacts in the $(2,2)$ mode waveform; and SXS:BBH:1111 lacks a waveform extracted at the outermost radius.

\item Lastly, we exclude 17 simulations in which the remnant spin flips to align opposite to the orbital angular momentum, i.e., $\thetaf \simeq \pi$ (see below for our parameter conventions). This is a known phenomenon in BH binary dynamics~\cite{Apostolatos:1994mx,Zhao:2017tro,Lousto:2014ida,Gerosa:2018mwg}, but it is difficult to model due to the limited number of such simulations available.
\end{itemize}

Following these cuts, our training set consists of 1866 NR simulations of which 61 are non-spinning, 423 have aligned spins, and 1382 have precessing spins. Throughout this paper, we use the following classification criteria:
\begin{enumerate}
\item  \textit{non-spinning} systems, where all spin components are below $10^{-4}$; 
 \item \textit{spin-aligned} systems, where all in-plane spin components $(\chi_{1x,y}, \chi_{2x,y})$ are $<10^{-4}$; 
 \item \textit{precessing} systems, where at least one in-plane spin component exceeds $10^{-4}$. 
 \end{enumerate}
 Non-spinning systems are a subset of spin-aligned systems, and unless specified, we treat them as part of the spin-aligned group.

For each SXS simulation, binary parameters such as BH masses and spins are measured at a reference time after the system reaches quasi-equilibrium. In the catalog, all quantities are provided in a coordinate frame where the $z$-axis is aligned with the instantaneous orbital angular momentum at the start of the simulation. In spin-aligned binary BHs, the orbital angular momentum direction remains constant (this is not true for binaries in the so-called up-down instability~\cite{Gerosa:2015hba,Varma:2020bon}, which however develops on much longer timescales compared to those considered here). On the other hand, in precessing systems, the orbital angular momentum precesses around the total angular momentum, and the BH spins evolve due to spin-spin and spin-orbit couplings. %
For each simulation, we measure time-dependent quantities relative to the inspiraling BHs at the time when the separation reaches the prograde innermost stable circular orbit (ISCO)~\cite{Bardeen:1972fi} of the remnant, which serves as our reference point throughout the paper. SXS simulations are provided with instantaneous positions and velocities of the BH apparent horizons in the starting coordinate frame. We compute the orbital angular momentum vector at ISCO using the classical definition $\boldsymbol{L}^{\rm ISCO} = M_1(\boldsymbol{r}^{\rm ISCO}_1 \times \boldsymbol{v}^{\rm ISCO}_1) + M_2(\boldsymbol{r}^{\rm ISCO}_2 \times \boldsymbol{v}^{\rm ISCO}_2)$.
QNM calculations, which determine frequencies and damping times, are performed within perturbation theory. In this framework, the natural choice is to work in a coordinate system whose $z$-axis is aligned with the spin direction of the remnant BH, $\chif$, a.k.a the ringdown frame. We compute the angle $\thetaf = \cos^{-1}{(\boldsymbol{L}^{\rm ISCO} \cdot \chif)}$ and apply a Wigner rotation to the waveform and associated vectorial quantities, transforming them from the initial frame—where the $z$-axis is aligned with the orbital angular momentum at the start of the simulation—to the ringdown frame using the publicly available SXS code package~\cite{Boyle:2019kee}.
Consequently, all amplitude fits and surrogate models are provided in the ringdown frame.

\begin{figure*}
    \centering
    \includegraphics[width=0.37\textwidth]{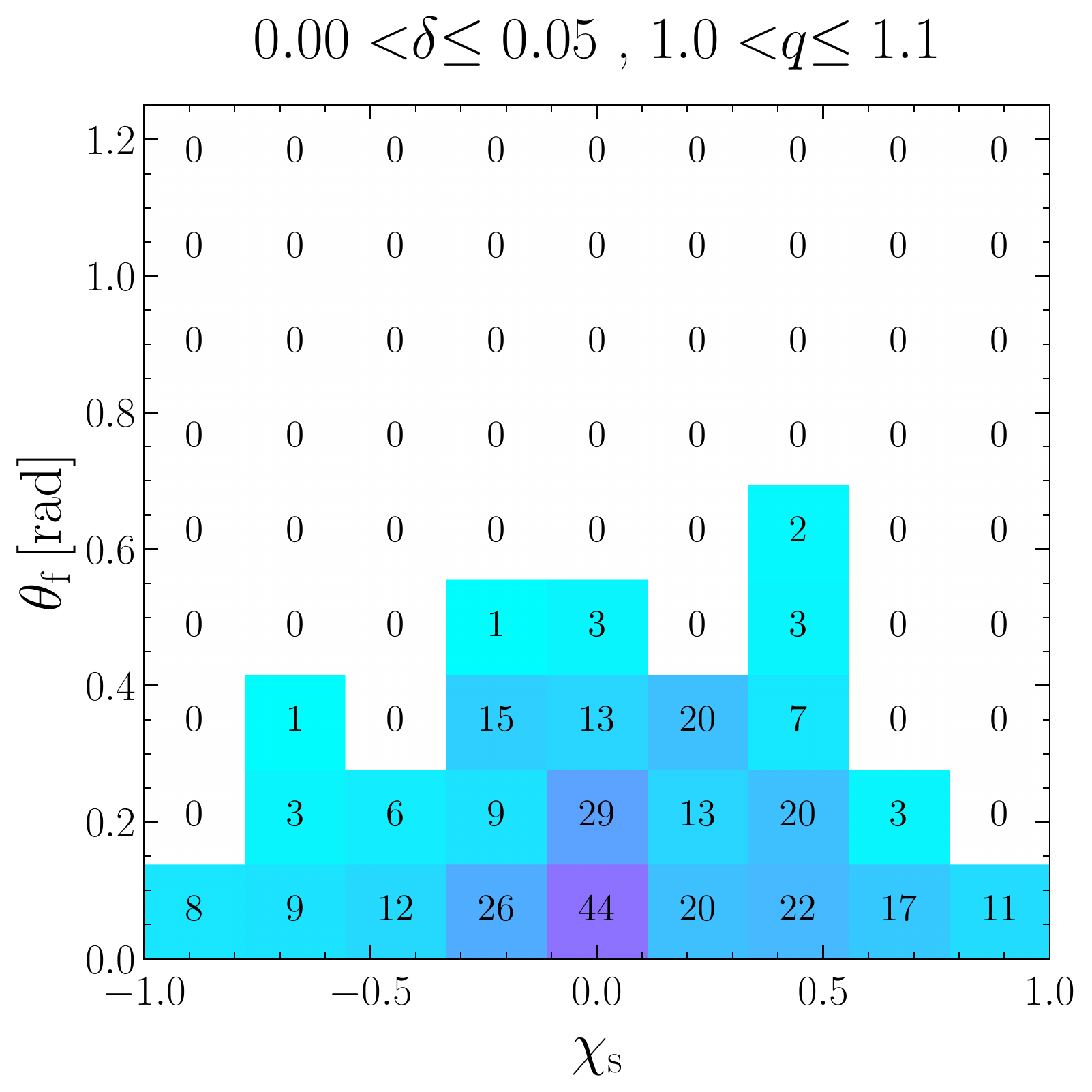}
    $\qquad\quad$
    \includegraphics[width=0.37\textwidth]{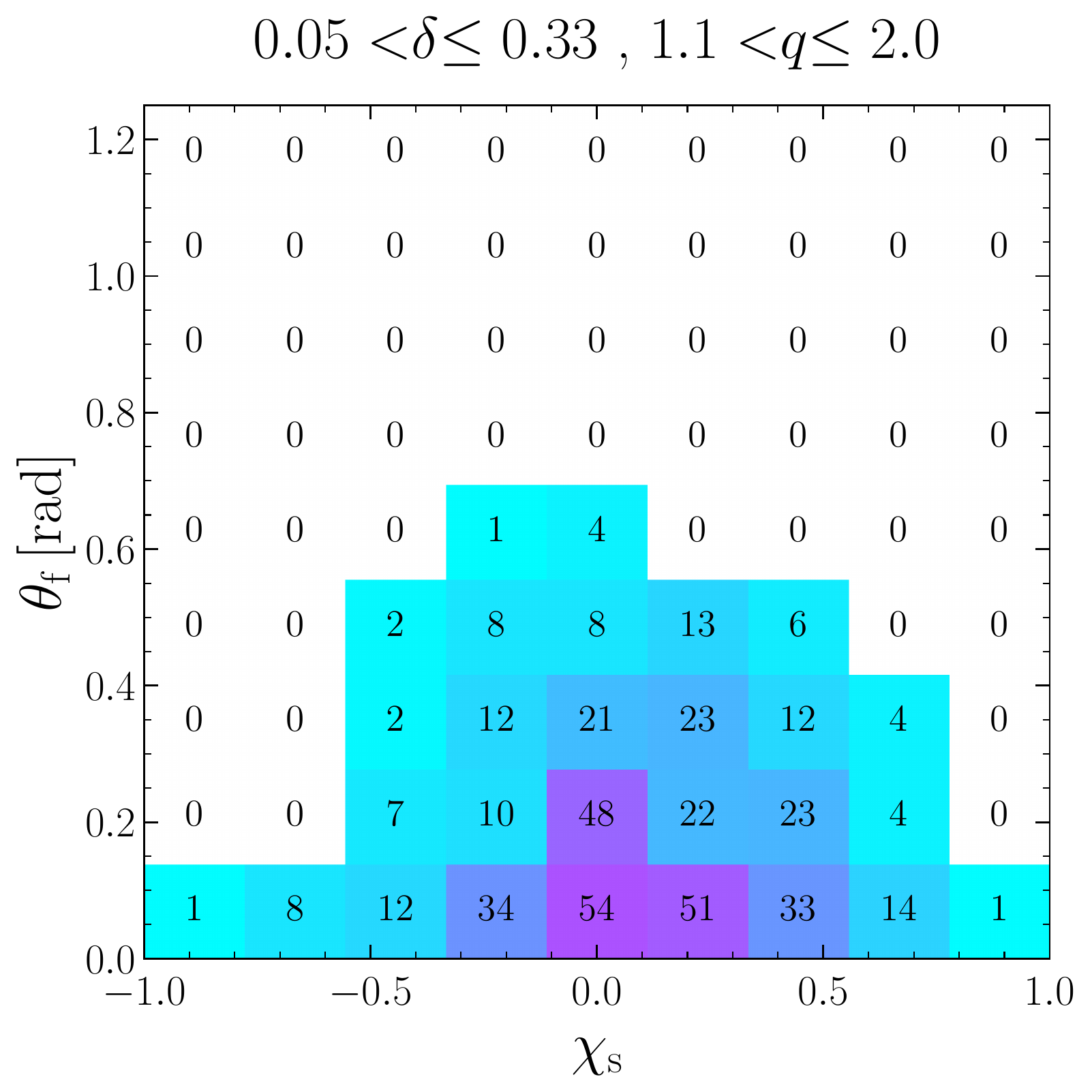} 
       \\  \bigskip
    \includegraphics[width=0.37\textwidth]{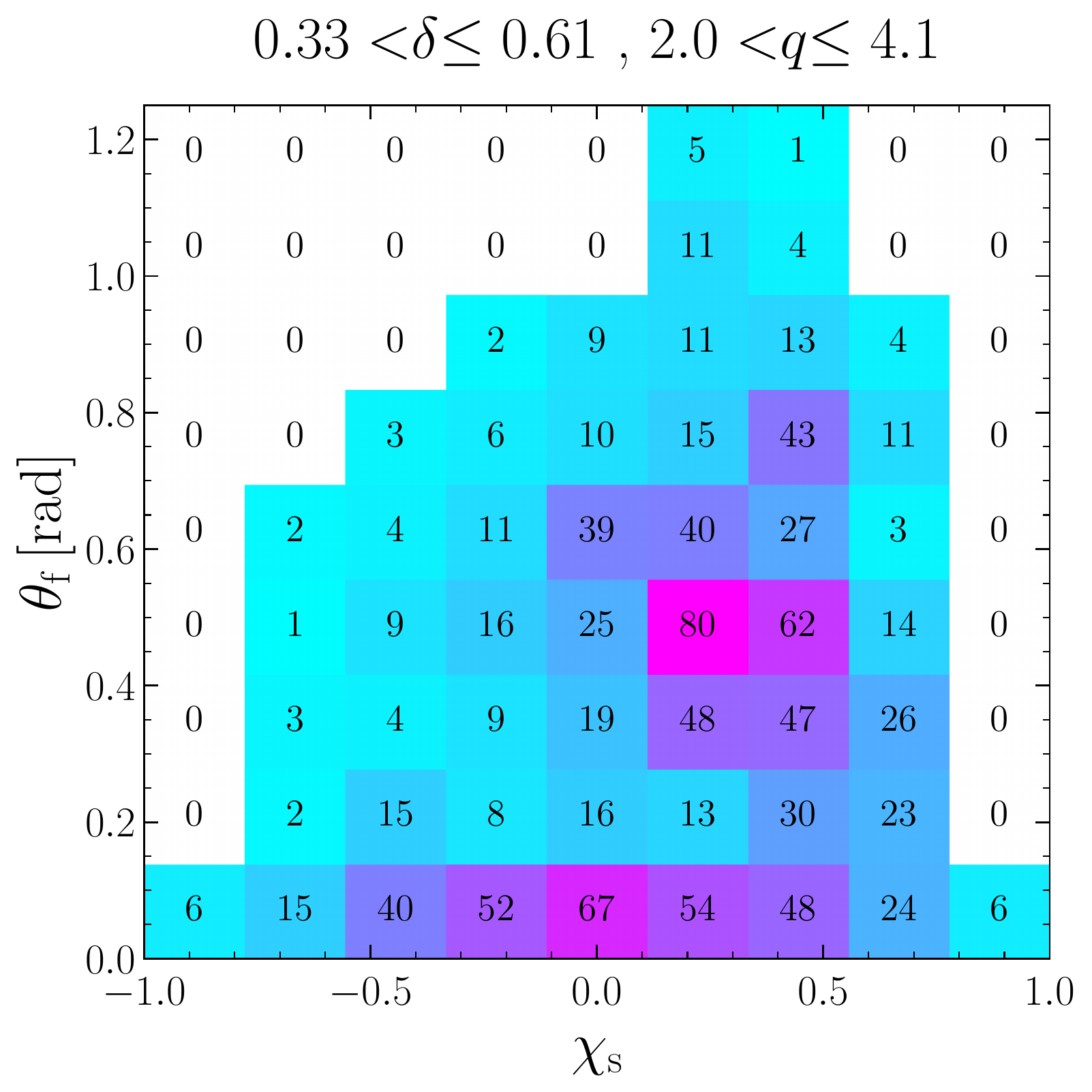} 
    $\qquad\quad$
    \includegraphics[width=0.37\textwidth]{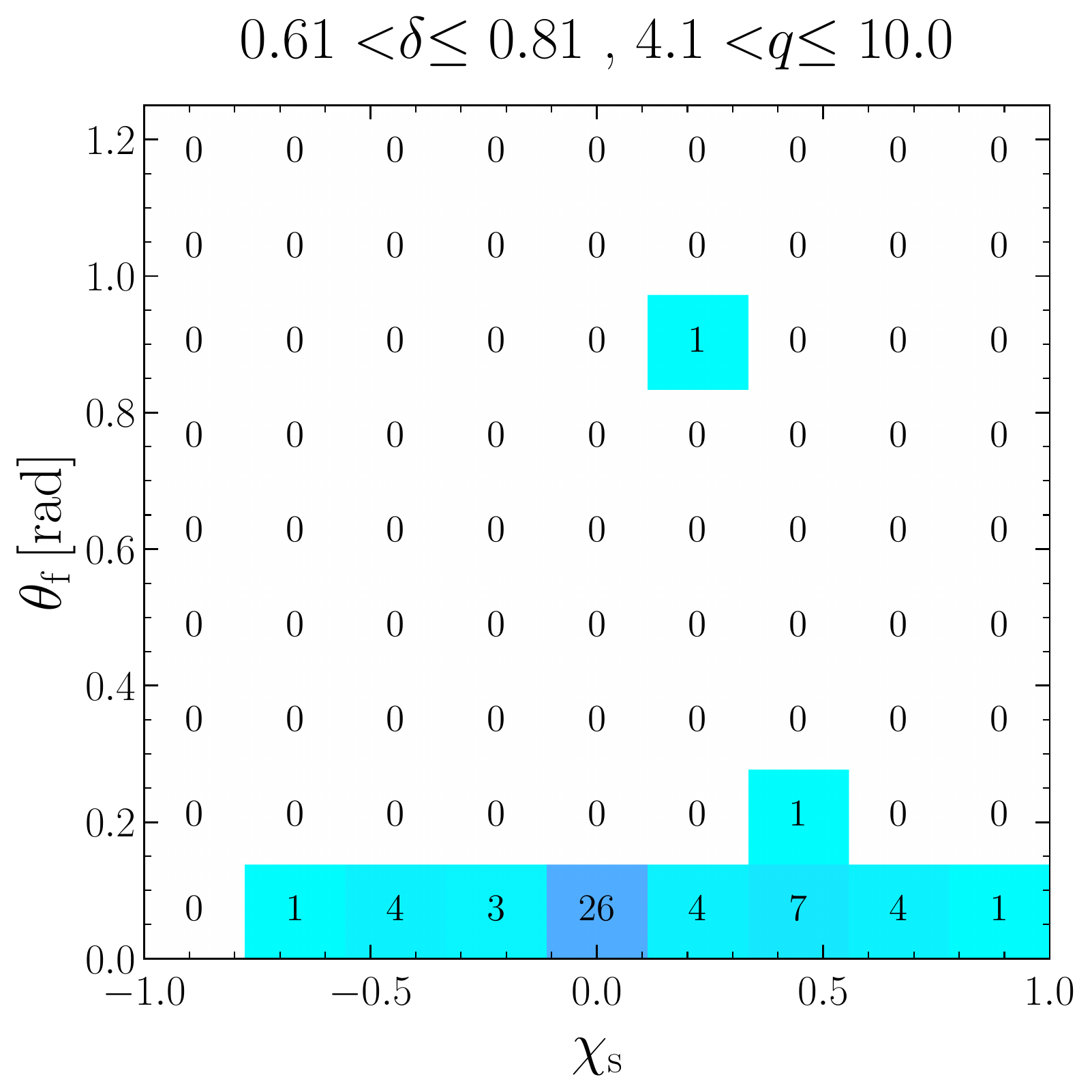}
    \caption{Distribution of SXS simulations in the ${(\delta, \chis, \thetaf)}$ parameter space. Each panel refers to a different mass ratio intervals; numbers and colors indicate the number of simulations in each bin.}
    \label{fig:param_space_2dhist}
\end{figure*}

\subsection{Parametrizations}
\label{sec:parameter_space}

When performing regression from binary BH parameters to mode amplitudes, identifying the optimal combination of parameters that effectively and efficiently captures amplitude features is non-trivial and may vary across different modes. However, identifying a suitable parametrization is crucial for accurate ringdown modeling. By experimenting with various combinations of physically motivated parameters of the binary BH system, we find that the $(2,m)$ and $(3,\pm 3)$ mode amplitudes are well described, and thus fit more effectively, when using the following six parameters: the asymmetric mass ratio $\delta = (q-1)/(q+1)$, with $q=m_1/m_2\geq 1$; the symmetric and antisymmetric combinations of the spin components parallel to the orbital angular momentum $\chi_{\rm s} = (q\chi_{1z} + \chi_{2z})/(1+q)$ and $\chi_{\rm a} = (q\chi_{1z} - \chi_{2z})/(1+q)$  (these are  often referred to as $\chi_{\rm eff}$ and $\delta\chi$ in the GW literature), %
the angle $\thetaf$ between the orbital angular momentum at ISCO and the remnant spin vector $\chif$, the angle $\phik$ between $\chif$ and the direction of the recoil, %
and the magnitude of recoil velocity $v_{\rm k}$.

Our guiding principle behind the parameter choices to capture the effect of precession is that the final spin encodes details on the precession dynamics before the merger~\cite{Kesden:2010yp,Varma:2018aht}, while the remnant kick is a good predictor of the asymmetry between $m>0$ and $m<0$ modes~\cite{Leong:2025raf, Borchers:2022pah, Mielke:2024kya}. Further, our choices are also inspired by insights of the effects of precession studied in Ref.~\cite{Hamilton:2021pkf, Thompson:2023ase} in the context of waveform modeling. Finally, a very similar choice of parameterization to ours was also recently explored for ringdown amplitudes in~\cite{Zhu:2023fnf}. %

Figure~\ref{fig:param_space_2dhist} shows the distribution of SXS simulations in the $(\delta, \chis, \thetaf)$ parameter sub-space. Most simulations fall within the range $0.05 < \delta < 0.6$, while only a few extend to higher mass ratio. Additionally, the vast majority of systems with $\thetaf \gtrsim 0.2$ exhibit positive values of $\chis$. Notably, all simulations with $\thetaf \gtrsim 0.7$ correspond to large mass ratios $\delta > 0.4$. 
We also observed that the dataset lacks precessing simulations in the region where $\thetaf \gtrsim 0.1$ and $|\chis| \gtrsim 0.7$.

Overall, for the modes we focus on in this study, we observe smooth variations in the ringdown amplitudes across this 6-dimensional (6D) parameter space. Each mode amplitude predominantly depends on two or three parameters, while the remaining ones have a subdominant effect. Although this parameter space is better suited for fitting and interpreting amplitude trends, it is not directly applicable to downstream GW applications -- particularly current data analysis pipelines typically require input in terms of Cartesian spin components. For this reason, we also develop a GPR fit in the 7-dimensional (7D) parameter space defined by $\delta$, $\vec{\chi_{1}}$, and $\vec{\chi_{2}}$ in addition to our 6-dimensional fits.  Note that this parameterization is less suitable for capturing amplitude features across the parameter space because of the lack of smooth variation in these parameters. While our regression may be less accurate for some modes in these parameters, we provide it for more practical and readily applicable predictions. Finally, possible alternative parameterizations are discussed in Appendix \ref{app:choice_param_space}.

\subsection{Ringdown fitting}
\label{sec:amp_extraction}

Extracting QNMs from NR simulations involves identifying the onset of the ringdown phase, which is still a challenging problem due to its ill-defined nature~\cite{Andersson:1996cm,Nollert:1998ys}. For spin-aligned systems, a common choice is to define the start of ringdown as $t_0 = t_{\rm peak} + \Delta t$, where $t_{\rm peak}$ is the peak of the $(2,2)$ strain mode. Conservative choices like $\Delta t \simeq 20M$ ensure that the system is well within the perturbative regime  and are favored choice for ringdown modeling~~\cite{Kamaretsos:2012bs, London:2014cma, London:2018gaq, Baibhav:2023clw, Pacilio:2024tdl}, 
whereas smaller values are sometimes chosen to maximize the signal-to-noise ratio (SNR), albeit at the risk of introducing modeling biases~\cite{Capano:2021etf, LIGOScientific:2021sio, LIGOScientific:2020tif, Forteza:2022tgq}. Alternative approaches based on physical intuition or fit stability have also been proposed~\cite{Cheung:2022rbm, Bhagwat:2017tkm,Bhagwat:2019dtm,Price:2023ldu}.To balance modeling biases against SNR loss, QNM amplitude models are typically fitted at conservative ringdown start times (e.g., $\Delta t \simeq 20M$). The predicted amplitudes are then extrapolated to earlier times ($\Delta t \simeq 10$--$15M$) by applying a rescaling factor of $e^{(t_{\rm fit} - t_{\rm eval})/\tau_{lmn}}$. Previous studies, including Ref.~\cite{Carullo:2018sfu}, have shown that this procedure does not introduce significant biases in parameter estimation. We release our models with a built-in method to extrapolate amplitudes at previous times.
For systems with misaligned spins, using the peak of the $(2,2)$ mode as a reference is unreliable, as this mode may not be dominant and precession introduces long-timescale amplitude modulations. Instead, we adopt a slight variation of the traditional $t_{\rm EMOP}$—the time of maximum energy-maximized orthogonal projection (EMOP) across all $(2,m)$ modes~\cite{Berti:2007fi,Baibhav:2017jhs}—as our reference time. A detailed discussion and a GPR-based fit for estimating this reference time are provided in Appendix~\ref{app:t_emop}. Note that while we use EMOP to select the reference time for aligning our simulations, other options, such as those in Refs.~\cite{Hamilton:2021pkf, Zhu:2023fnf, Finch:2021iip},
exist, and the impact of reference time prescription on precessing ringdown models has not yet been studied. 
Moreover, note that $t_{\rm EMOP}$ does not reduce to $t_{\rm peak}$ in the non-precessing limit, as clearly visible from Fig.~\ref{fig:t_emop_scatter} in Appendix~\ref{app:t_emop}.
Adopting a conservative approach, we define the ringdown starting time as $t_0 = t_{\rm EMOP} + 20M$. Although this choice remains somewhat arbitrary, we find that the fit error $\epsNR$ (see below) typically plateaus around this time, providing an empirical validation for this choice.

After isolating the ringdown, we fit the $(2,m)$ and $(3,\pm3)$ modes independently for each SXS simulation. Additionally, the real and imaginary components of the $(2,0)$ mode are treated as independent modes, since $m=0$ modes are not circularly polarized (see Sec.~\ref{subsec:20_mode}). Each mode is fitted using a least-squares algorithm implemented via \textsc{scipy.minimize}~\cite{Virtanen:2019joe}, with the following ansatz:
\begin{equation}
\label{eq:waveform}
    h_{lm}(t) = A_{lm} \,e^{-(t-t_0)/\tau_{lm}} e^{-i\omega_{lm}(t-t_0) + \phi_{lm}} 
\end{equation}
The amplitudes $A_{lm}$ and phases $\phi_{lm}$ are the free parameters of out fits. The QNM frequencies $\omega_{lm}$ and damping times $\tau_{lm}$ are fixed using the tables\footnote{See Ref.~\cite{Pacilio:2024tdl} for clarifications regarding the different notation conventions adopted in Ref.~\cite{Berti:2005ys}.} of Ref.~\cite{Berti:2005ys}, which we interpolated. We truncate all waveforms at $t=t_0+100M$ to exclude spurious effects due to numerical noise.

Note that there are several simplifying assumptions inherent to this ansatz. 
\begin{enumerate}
\item
First, it does not account for spherical-spheroidal mode mixing. This effect is typically small for the modes we consider, especially when the spin of the remnant is moderate~\cite{Berti:2014fga,London:2014cma}. For detailed modeling of mixing coefficients see Ref.~\cite{Berti:2014fga,London:2018nxs,Cook:2014cta,Stein:2019mop}; %
we leave their implementation to future exploration.
\item 
Second, we assume that all overtones have sufficiently decayed by $t_0$, which is reasonable given that overtone half-lives are typically shorter than half a wave cycle~\cite{Bhagwat:2019dtm,Gennari:2023gmx}. For example, in a near-equal mass system, the damping time of the first overtone ($n=1$) is about a quarter of that of the dominant mode ($n=0$).
\item Third, we do not include the excitation of nonlinear modes; while these can be important for higher-$l$ modes such as $(4,4)$, they remain negligible for $(2,m)$ and $(3,\pm3)$ as first studied in Ref.~\cite{London:2014cma} and more recently in Refs.~\cite{Mitman:2022qdl, Cheung:2022rbm, Cheung:2023vki}. 
\item Finally, we choose to neglect the excitation of retrograde modes. We explicitly checked the amplitudes of retrograde modes in our simulation set and verify that it is a good approximation for the parameter space of our simulations. Retrograde modes are significantly excited in cases where the system's angular momentum reverses during evolution~\cite{Apostolatos:1994mx,Zhao:2017tro,Lousto:2014ida,Gerosa:2018mwg}, a phenomenon we do not capture due to the sparsity of our training set (see Sec.~\ref{sec:conventions}). Further investigations on retrograde modes are presented in Appendix \ref{app:retrograde_modes}.
\end{enumerate}

The SXS code developers recommend a threshold of $h = 10^{-5}$ beyond which numerical errors dominate~\cite{Boyle:2019kee}. In our fitting process, we adopt a more cautious threshold of $h = 10^{-4}$, below which we treat all amplitudes as zero.

We evaluate the quality of our least-square waveform fits using the relative error
\begin{equation}
\label{eq:eps_nr}
    \epsNR = \frac{\int_{t_0}^{100M}\left| h_{lm}^{\rm NR} - h_{lm}^{\rm fit} \right|^2 dt}{\int_{t_0}^{100M}\left| h_{lm}^{\rm NR} \right|^2 dt},
\end{equation}
where $h_{lm}^{\rm fit}$ is the QNM model of Eq.~\eqref{eq:waveform} with amplitude and phase determined by the fitting procedure and $h_{lm}^{\rm NR}$ is the NR simulation. %

\subsection{Mode amplitude distribution}
\label{sec:amp_extr_results}

\begin{figure}
     \centering
     \includegraphics[width=\columnwidth]{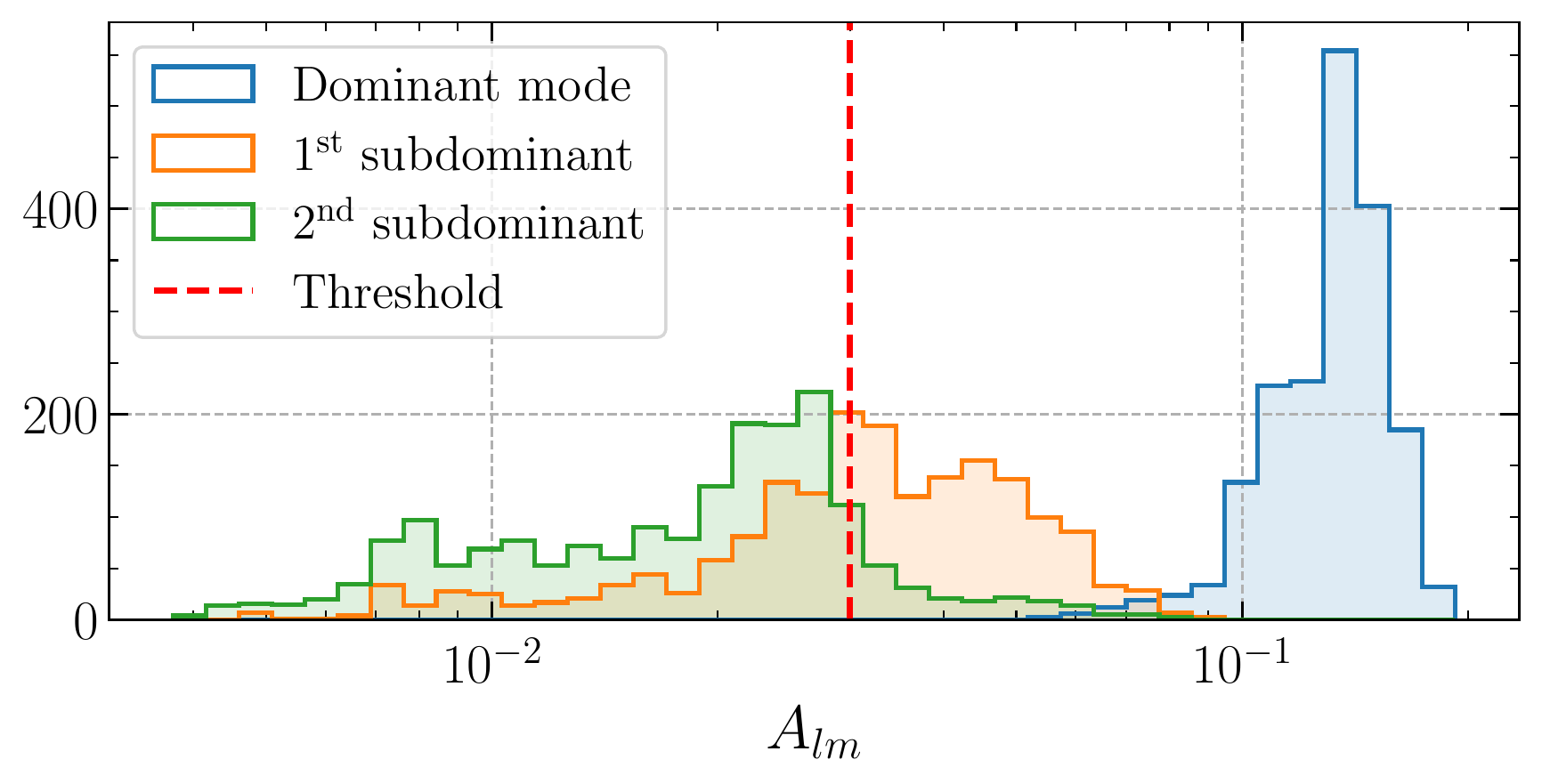}
     \caption{Distribution of ringdown amplitudes for the dominant and first two subdominant modes. The nominal threshold $A_{lm} = 0.03$ helps identify modes expected to play a significant role.}
     \label{fig:dominant_amp_distrib_hist}
 \end{figure}

Figure~\ref{fig:dominant_amp_distrib_hist} shows the distributions of the fitted amplitude values of the dominant and first two subdominant modes in the data set. Here we do not identify specific modes, but rather rank them by amplitude and compare the corresponding excitation levels. For this figure, we do not distinguish between $m > 0$ and $m < 0$ modes, as their amplitudes produce nearly identical distributions. Instead, we separate modes depending on their frequencies. These distributions provide useful information about the overall amplitude scale and the relative factors between dominant and subdominant modes, even before modeling correlations with physical parameters of the sources.

 \begin{figure*}[ht]
    \centering
    \includegraphics[width=2\columnwidth]{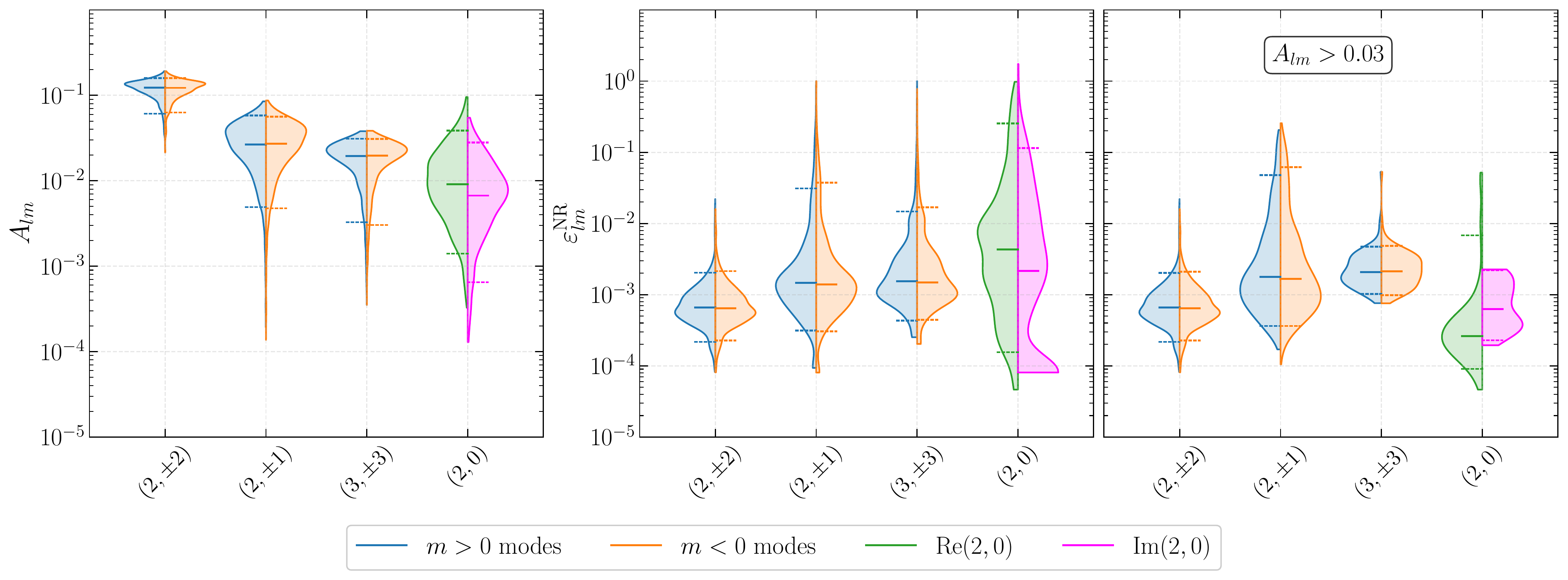}
    \caption{Distributions of fitted amplitude values $A_{lm}$ (left panel),  fit errors $\epsNR$ (middle panel), and fit errors related to simulations with $A_{lm}>10^{-2}$ (right panel). Solid lines correspond to median values, while dashed lines refer to $5^{\rm th}$ and $95^{\rm th}$ percentiles.
    }
    \label{fig:fits_3panel}
\end{figure*}

Across the parameter space, the amplitude of the dominant mode typically falls within $0.05 < A_{lm} < 0.2$, while the first two subdominant modes generally satisfy $A_{lm} < 0.1$. A few simulations exhibit comparable excitation levels between dominant and subdominant modes. We explore mode hierarchy in more detail in Sec.~\ref{sec:mode_hierarchy}.
Based on Fig.~\ref{fig:dominant_amp_distrib_hist}, we set a heuristic threshold at $A_{lm} > 0.03$. This value is close to the peak of the amplitude distribution for the first subdominant mode and ensures that at least $\sim 100$ simulations are included in this subsample for the $(3,3)$, $\text{Re}(2,0)$, and $\text{Im}(2,0)$ modes. We tag these simulations as ``significantly excited'' and use this classification to interpret our results.

Results from our ringdown-extraction procedure are illustrated in Fig.~\ref{fig:fits_3panel}, where we show the distributions of the fitted amplitude values $A_{lm}$, fit errors $\epsNR$, and fit errors for simulations with $A_{lm} > 0.03$.
We observe that $\epsNR$ for the $(2,\pm2)$, $(2,\pm1)$, and $(3,\pm3)$ modes generally falls within the range $10^{-4}$ to $10^{-2}$. However, errors for the $(2,0)$ mode can be significantly higher. Notably, when considering only systems where $A_{20} > 0.03$, the fit error for the $(2,0)$ mode falls within the same range as the other modes.

\section{Phenomenology}
\label{sec:amp_phenom}

\begin{table*}[t]
\centering
\renewcommand{\arraystretch}{1.3} %
\setlength{\tabcolsep}{2pt} %
\resizebox{\textwidth}{!}{%
\begin{tabular}{>{\centering\arraybackslash}m{2cm}||>{\centering\arraybackslash}m{3cm}|>{\centering\arraybackslash}m{3cm}|>{\centering\arraybackslash}m{3cm}|>{\centering\arraybackslash}m{3cm}|>{\centering\arraybackslash}m{3cm}|>{\centering\arraybackslash}m{3cm}}

  {Mode} &
  {$\delta$} &
  {$\thetaf$} &
  {$\chis$} &
  {$\chia$} &
  {$\phik$} &
  {$v_{\rm k}$} \\ \hline \hline
{(2,2)} &
  \hlc[red!20]{Negative, strong} &
  \hlc[red!20]{Negative, strong}  &
  {-} &
  {-} &
  \hlc[cyan!20]{Positive, strong} &
  {\begin{tabular}[c]{@{}c@{}}$\phik \sim \pi$ :\\ \hlc[cyan!20]{Positive, strong}\\ $\phik \sim 0$ :\\ \hlc[red!20]{Negative, strong}\end{tabular}} \\ \hline
{(2,-2)} &
  \hlc[red!20]{Negative, strong}  &
  \hlc[red!20]{Negative, strong}  &
  {-} &
  {-} &
  \hlc[red!20]{Negative, strong}  &
  {\begin{tabular}[c]{@{}c@{}}$\phik \sim \pi$ : \\\hlc[red!20]{Negative, strong}\\ $\phik \sim 0$ : \\\hlc[cyan!20]{Positive, strong}\end{tabular}} \\ \hline
{(2,1)} &
  {Negative, weak} &
  \hlc[cyan!20]{Positive, strong} &
  \hlc[red!20]{Negative, strong}  &
  \hlc[red!20]{Negative, strong}  &
  {-} &
  {$\phik \sim \frac{\pi}{2}$: Positive} \\ \hline
{(2,-1)} &
  {Negative, weak} &
  \hlc[cyan!20]{Positive, strong} &
  \hlc[red!20]{Negative, strong}  &
  \hlc[red!20]{Negative, strong}  &
  {-} &
  {$\phik \sim \frac{\pi}{2}$: Positive} \\ \hline
{(3,3)} &
  \hlc[cyan!20]{Positive, strong} &
  {\begin{tabular}[c]{@{}c@{}}for large $\delta$: \\\hlc[red!20]{Negative, strong}\end{tabular}} &
  {for large $\delta$: Positive} &
  {for large $\delta$: Positive} &
  {for large $\delta$: Positive} &
  {-} \\ \hline
{(3,-3)} &
  \hlc[cyan!20]{Positive, strong} &
{\begin{tabular}[c]{@{}c@{}}for large $\delta$: \\\hlc[red!20]{Negative, strong}\end{tabular}} &
  {for large $\delta$: Positive} &
  {for large $\delta$: Positive} &
  {for large $\delta$: Negative} &
  {-} \\ \hline
  Re(2,0) &
  {\begin{tabular}[c]{@{}c@{}}for $\chis \rightarrow -0.8$: \\Positive, weak \end{tabular}}&
  \hlc[cyan!20]{Positive, strong} &
  Negative &
  Negative, weak &
  -&
  - \\ \hline
  Im(2,0) &
  {\begin{tabular}[c]{@{}c@{}}for $\chis \rightarrow -0.8$: \\Positive, weak \end{tabular}} &
  \hlc[cyan!20]{Positive, strong} &
  \hlc[red!20]{Negative, strong} &
  \hlc[red!20]{Negative, strong} &
  - &
  - 
\end{tabular}%

}
\caption{Correlations between mode amplitude and binary parameters, using the 6D parametrization described in Sec.~\ref{sec:parameter_space}. We distinguish between positive and negative correlations, as well as strong and weak ones (this is judged qualitatively by eye). Hyphens indicate parameters for which we do not observe a clear correlation in any direction. For ease of reading, strong negative correlations are highlighted in peach-pink, while strong positive correlations are shown in sky-blue.
}
\label{tab:trend_summary}
\end{table*}

In this section, we examine the variations in QNM amplitudes for all systems in our dataset across the 6D parameter space $\left( \delta, \chis, \chia, \thetaf, \phik, v_{\rm k} \right)$. Identifying clear patterns is challenging since amplitude variations do not follow a simple hierarchical dependence on parameters. Some modes are more sensitive to specific parameters, while others have a weaker, subdominant influence on their overall variation. 

The most significant correlations are summarized in Table~\ref{tab:trend_summary}. In the following subsections, we present figures showcasing the amplitude variations along selected key parameters on the $x$- and $y$-axes, with amplitude values represented on the color scale. %
Since we are projecting 6D parameter dependencies onto 2D plots, the resulting patterns may appear less smooth, reflecting the need for a nuanced, multidimensional parameterization to capture the behavior of amplitudes.
For simplicity, in some of the plots below, we only show positive-$m$ modes, but the qualitative behavior we highlight also applies to negative-$m$ modes, with a few differences, which are listed in Table~\ref{tab:trend_summary}. %

\subsection{Amplitudes of the $(2,\pm2)$ modes}
\label{sec:22_mode_amp}

A large $\thetaf$ angle is typically associated with systems that have a large value of $\delta$. The amplitudes of the $(2, \pm 2)$ modes decrease as both $\delta$ and $\thetaf$ decrease, as shown in the top-left panel of Fig.~\ref{fig:2m-mode}.

The top-right panel of Fig.~\ref{fig:2m-mode} shows the relationship between the $(2,2)$ mode amplitude and the kick-spin angle $\phik$. Both $\phik$ and $v_{\rm k}$ strongly influence the amplitude of the $(2, \pm 2)$ modes, and these parameters primarily determine the breaking of equatorial symmetry between the $(2,2)$ and $(2,-2)$ mode amplitudes~\cite{Leong:2025raf, Borchers:2022pah, Mielke:2024kya}. Notably, higher $v_{\rm k}$ values typically occur near $\phik \simeq 0$ or $\phik \simeq \pi$~\cite{Gerosa:2018qay}. Unlike the amplitude variations seen with respect to $\delta$ and $\thetaf$, where both modes exhibit similar responses, their behavior with respect to changes in $\phik$ and $v_{\rm k}$ is opposite. For the $(2,2)$ mode, as $\phik \rightarrow \pi$ and $v_{\rm k}$ increases, the amplitude increases. In contrast, when $\phik \sim 0$, a higher value of $v_{\rm k}$ leads to a decrease in amplitude. The $(2,-2)$ mode amplitude shows the opposite behavior to that seen for $(2,2)$ : as $\phik \rightarrow \pi$ and $v_{\rm k}$ increases, the amplitude decreases; when $\phik \sim 0$, a higher value of $v_{\rm k}$ leads to an increase in amplitude.

Additionally, we note that the amplitude variations of the $(2, \pm 2)$ modes with respect to the aligned spin parameters $\chis$ and $\chia$ are minimal.

\subsection{Amplitudes of the $(2,\pm1)$ modes}  

\begin{figure}
    \centering
    \includegraphics[width=\columnwidth]{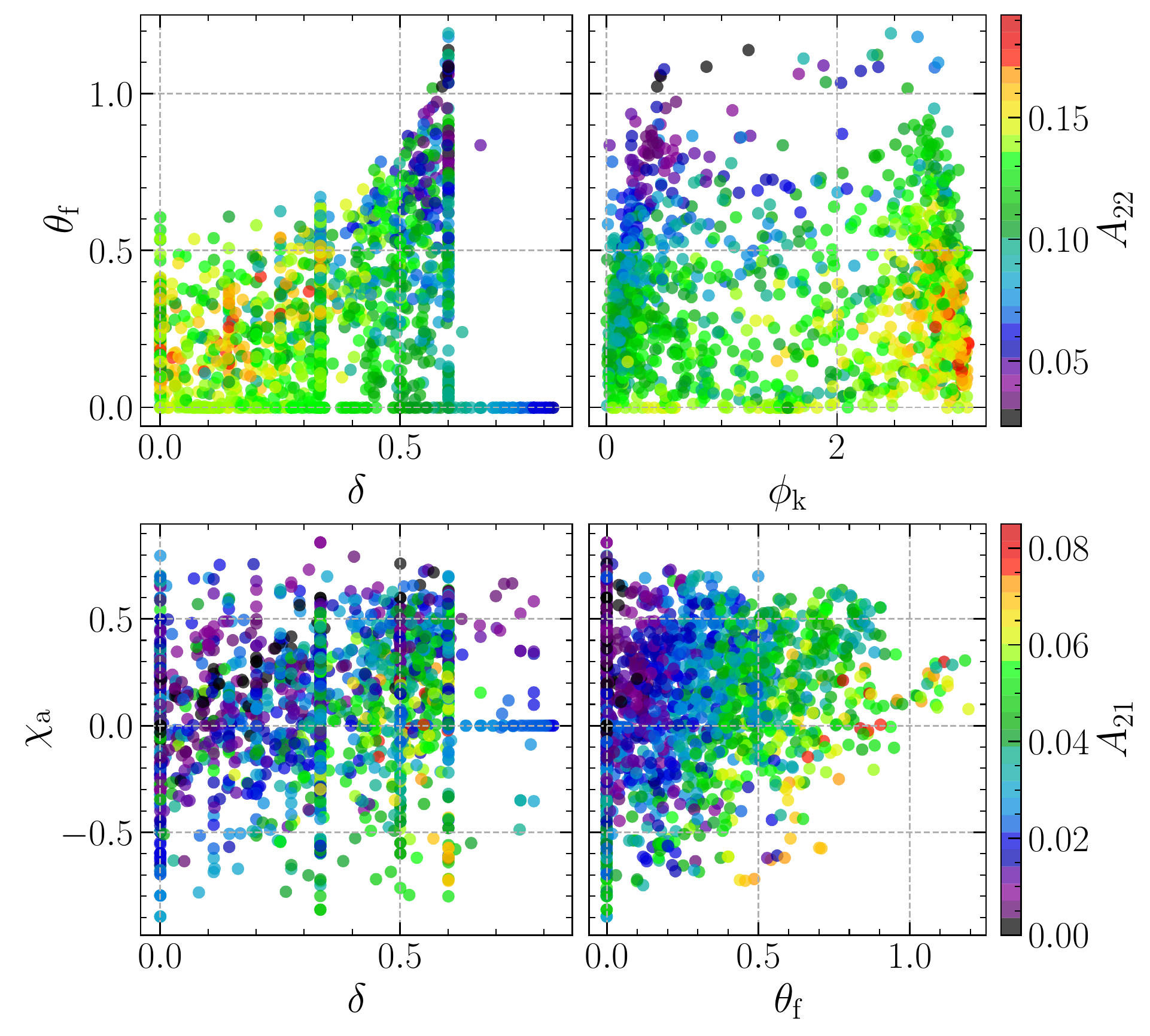}
    \caption{Amplitudes of the $(2,2)$ and $(2,1)$ modes across some 2D sections of our 6D parameter space. Each circle indicates a NR simulation.%
}
    \label{fig:2m-mode}
\end{figure}

In the bottom row of Fig.~\ref{fig:2m-mode}, we show notable correlations of the $(2,1)$ mode amplitude with respect to $\delta$, $\chia$, and $\thetaf$. The amplitudes of the $(2, \pm 1)$ modes are only weakly affected by the $\delta$ mass-ratio parameter, with a very slight increase as $\delta$ increases.
The most significant influences on the amplitude values of $(2, \pm 1)$ modes come from the opening angle $\thetaf$, and the aligned spin parameter $\chia$. As $\thetaf$ increases, the amplitudes also increase. The two aligned spin parameters show similar patterns, but $\chi_{\rm a}$ is characterized by a smoother trend: the amplitudes decrease as $\chi_{\rm a}$ increases, with the largest amplitude values occurring at large and negative values of $\chi_{\rm a}$. There is no clear correlation with $\phik$ in general, but near $\phik \simeq \pi/2$ and large $v_{\rm k}$, we observe a mild increase in the ringdown amplitude.

\subsection{Amplitudes of the $(3,\pm3)$ modes}

\begin{figure*}
    \centering
    \includegraphics[width=2\columnwidth]{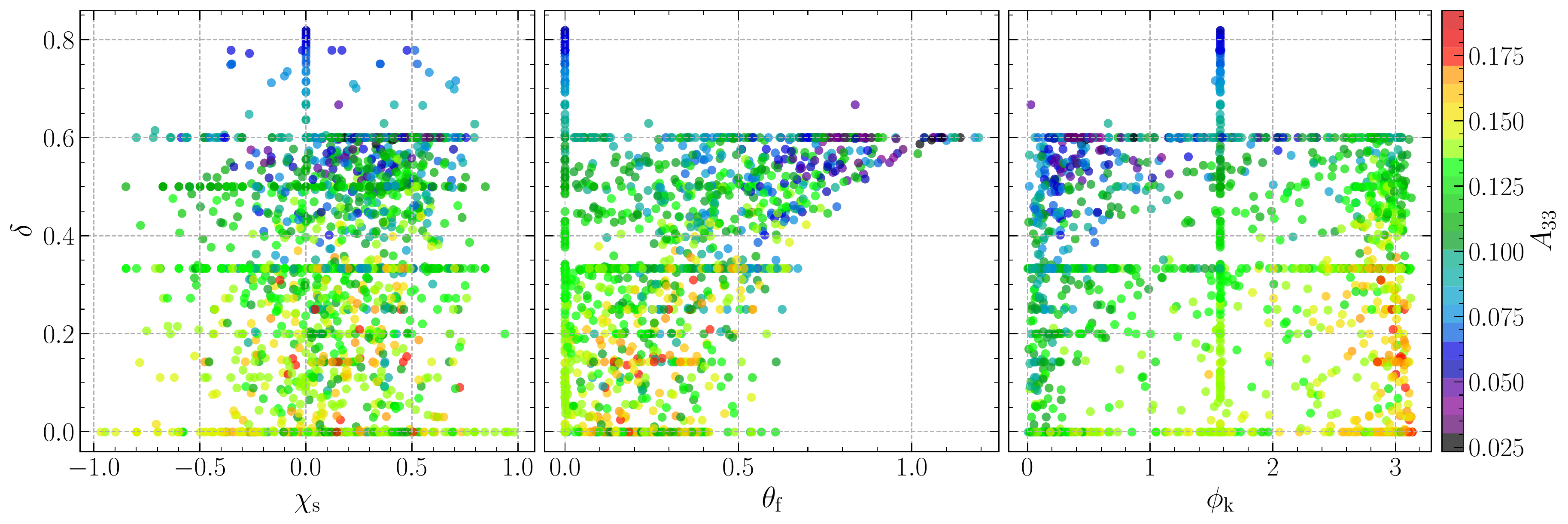}
    \caption{Amplitude of the $(3,3)$ more across some 2D sections of our 6D parameter space. Each circle indicates a NR simulation.}%
    \label{fig:33-mode}
\end{figure*}

The amplitudes of the $(3, \pm 3)$ modes exhibit less smooth variations, indicating that their values depend dominantly on more than two parameters, as shown in Fig.~\ref{fig:33-mode}. As $\delta$ increases, the amplitude of the $(3, \pm 3)$ modes decreases. Additionally, when $\thetaf \simeq 0$, the amplitude is primarily determined by $\delta$. We also observe that when both $\delta$ and $\thetaf$ are moderately small, the amplitudes tend to be larger. Figure~\ref{fig:33-mode} shows that there is no strong correlation with the spin-aligned parameters $\chi_{\rm s,a}$.
We do not observe significant dependencies in $\phik$ and $v_{\rm k}$, but equatorial symmetry breaking appears in the amplitudes of the $(3,3)$ and $(3,-3)$ modes (see Sec.~\ref{sec:mode_asymmetry}).

\subsection{Amplitudes of the $(2,0)$ mode}
\label{subsec:20_mode}
\begin{figure}
    \centering
    \includegraphics[width=\columnwidth]{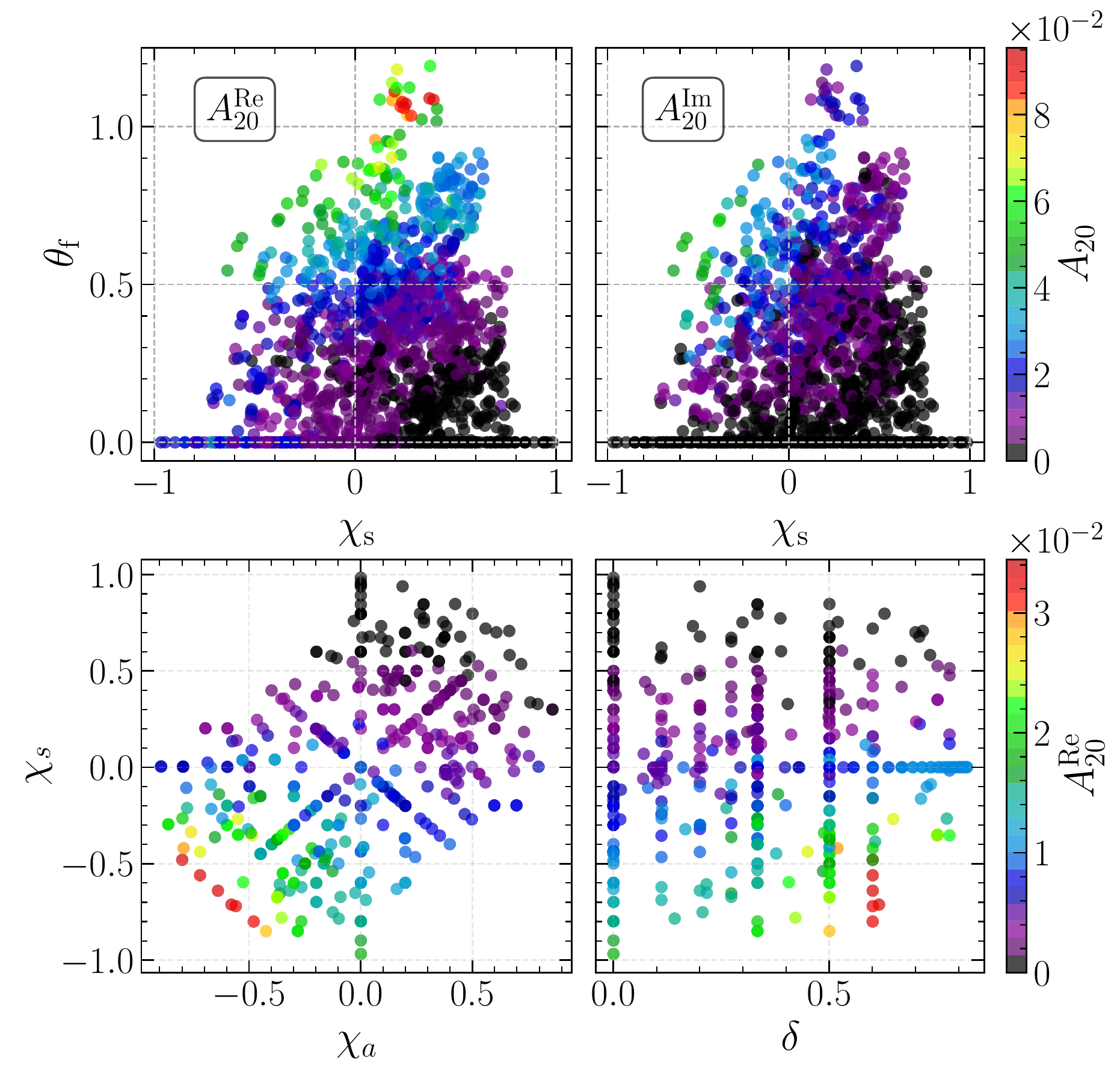}
    \caption{
    Amplitude of the real and imaginary part of the $(2,0)$ mode across some 2D sections of our 6D parameter space. Each circle indicates a NR simulation. The top panels include both spin-aligned and precessing systems, while the bottom panel focuses exclusively on spin-aligned systems. Notably, we observe a significant increase in the $(2,0)$ mode amplitude for negative values of $\chi_{\rm s}$ and $\chi_{\rm a}$, even in the absence of precession.}
    \label{fig:20-mode}
\end{figure}

The $m=0$ modes are not circularly polarized, so we model the real and imaginary parts of the waveforms independently, extracting two separate amplitude values: $A^{\rm Re}_{20}$ and $A^{\rm Im}_{20}$. In Fig.~\ref{fig:20-mode}, we show the dominant correlations in the parameter space, separating precessing and aligned systems. The primary parameters influencing the amplitude of the $(2,0)$ mode are the aligned spin component $\chi_{\rm a,s}$ and the angle $\thetaf$. 
Both the real and imaginary parts exhibit similar correlations: the ringdown amplitudes increase with $\thetaf$ and when $\chi_{\rm s,a}$ are both negative.

 \begin{figure*}
    \centering
    \includegraphics[width=1.9\columnwidth]{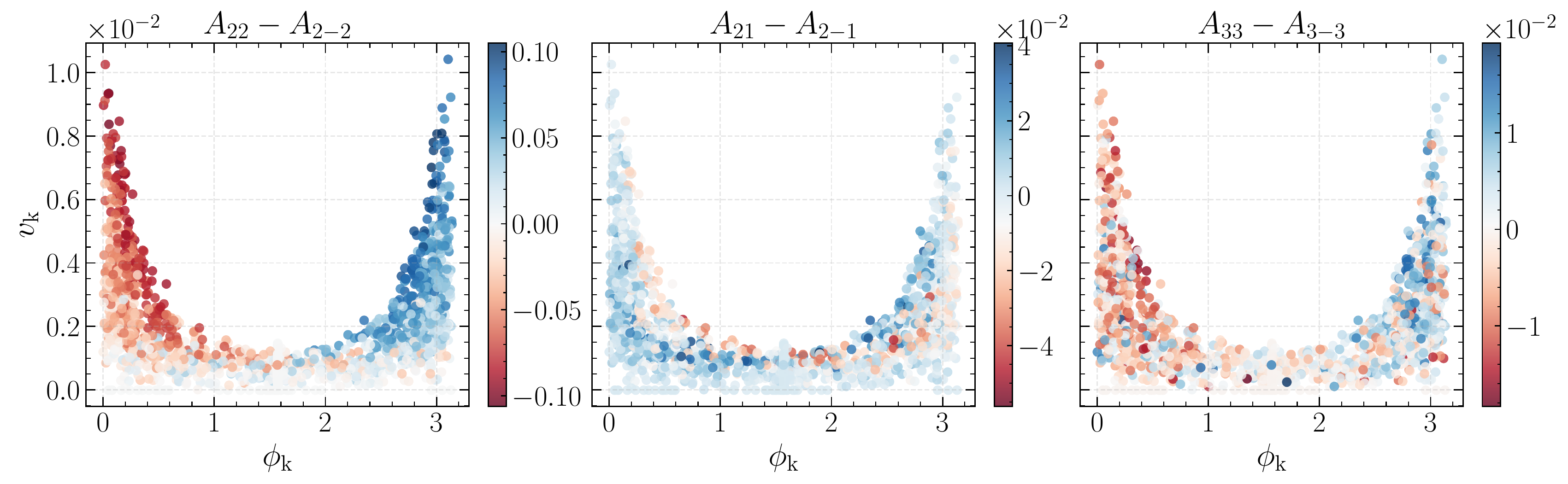}
    \caption{Mode asymmetry in BH ringdowns. We show the amplitude difference between $m>0$ and $m<0$ modes as a function of the kick direction $\phik$ and magnitude $v_{\rm k}$. Asymmetries are most pronounced when $\phik \simeq 0$ or $\phik \simeq \pi$.}
    \label{fig:equat_symm_scatterplots}
\end{figure*}

\subsection{Mode hierarchy}
\label{sec:mode_hierarchy}

For non-spinning systems, the excitation amplitudes follow a well-defined hierarchy: $A_{2\pm 2} \geq A_{3\pm 3} \geq A_{2\pm 1} \geq A_{44}$, and $A_{20} = 0$. This can be observed from the amplitude fits for the nonspinning systems presented in this work or in previous studies in the literature~\cite{London:2014cma,Bhagwat:2019bwv,Forteza:2022tgq,Pacilio:2024tdl,London:2018gaq,Cheung:2022rbm}. 

Similarly, from the fits for spin-aligned systems, we know that the $(2, \pm 2)$ modes remain dominant, but the first subdominant mode can be either $A_{3\pm 3}$ or $A_{2\pm 1}$, depending on the spin magnitudes and directions
In near-equal-mass systems with aligned spins, both BHs influence the amplitudes. The amplitude of the $(2, \pm 2)$ modes decreases with increasing $\chi_{\rm s}$, while the amplitudes of the $(2, \pm 1)$ and $(3, \pm 3)$ modes increase with increasing $|\chi_{\rm a}|$. Among the subdominant modes, the $(2, \pm 1)$ mode is typically the most significant, with its amplitude reaching about 20\% of the dominant $(2, \pm 2)$ mode. Additionally, a significant excitation of the $\rm{Re}(2, 0)$ mode is observed as $\chi_{\rm a,s}$ becomes more negative, while the $\rm{Im}(2, 0)$ mode remains unexcited.
In contrast, in the limit of high mass asymmetry for spin-aligned systems, where the spin of the larger black hole dominates, both the $(2, \pm 1)$ and $(3, \pm 3)$ modes can become significant. In this regime, the loudest subdominant mode typically has an amplitude between 30\% and 50\% of the $(2, \pm 2)$ mode. The amplitude of the $(2, \pm 2)$ mode decreases with increasing mass ratio $\delta$. Similarly, the amplitudes of the $(2, \pm 1)$ modes and the real part of $(2, 0)$ decrease with increasing $\chi_{\rm s}$, while the amplitude of the $(3, \pm 3)$ modes decreases with $\delta$ but increases with $\chi_{\rm s}$.

In precessing systems, the mode hierarchy is more complex and less predictable. While one of the $(2,m)$ modes generally remains dominant, the specific order can vary significantly depending on the parameters. In the case of equal masses, the $(2, \pm 2)$ mode remains consistently dominant. The first subdominant mode is typically $(2, \pm 1)$ (observed in 95\% of the sources in this subset), though $\rm{Re}(2,0)$ can occasionally become more prominent. The amplitude of the first subdominant mode generally falls between 5\% and 30\% of the dominant mode. In particular, the amplitudes of $(2, \pm 1)$, $(3, \pm 3)$, and $\text{Re}(2,0)$ modes tend to increase with $\thetaf$, although this variation is somewhat irregular. Additionally, the $(3,3)$ and $(3,-3)$ modes show a slight increase as $\phi_k \to \pi$.

Systems with large $\theta_{\rm f}$ angles, indicating strong precession, present an interesting phenomenology.
In such cases, the mode hierarchy becomes even more unpredictable, partly due to the limited number of simulations available. Depending on the parameters, any of the $(2,m)$ modes can dominate, though the $(2, \pm 2)$ mode still retains dominance most often. Among the 375 simulations in the SXS catalog with an  $\theta_{\rm f} >0.5$ %
 radians (approximately 30°), the $(2, \pm 2)$ mode is dominant in 365 cases. In 9 simulations, the dominant mode is $\text{Re}(2,0)$, and in 1 case, the dominant mode is $(2,1)$. The amplitude of the first subdominant mode can range from a few percent to nearly 100\% of that of the dominant mode. In particular, 121 simulations exhibit at least one subdominant mode with a ringdown amplitude that is at least $50\%$ of the dominant-mode amplitude. Of these, 73 simulations feature only the $(2, \pm 1)$ mode surpassing the threshold, while 43 simulations show both $(2, \pm 1)$ and $\text{Re}(2,0)$ above it. In 5 cases, the $(2, \pm 1)$, $\text{Re}(2,0)$, and $\text{Im}(2,0)$ modes all exceed the 50\% threshold.

 \begin{figure}
    \centering
    \includegraphics[width=\columnwidth]{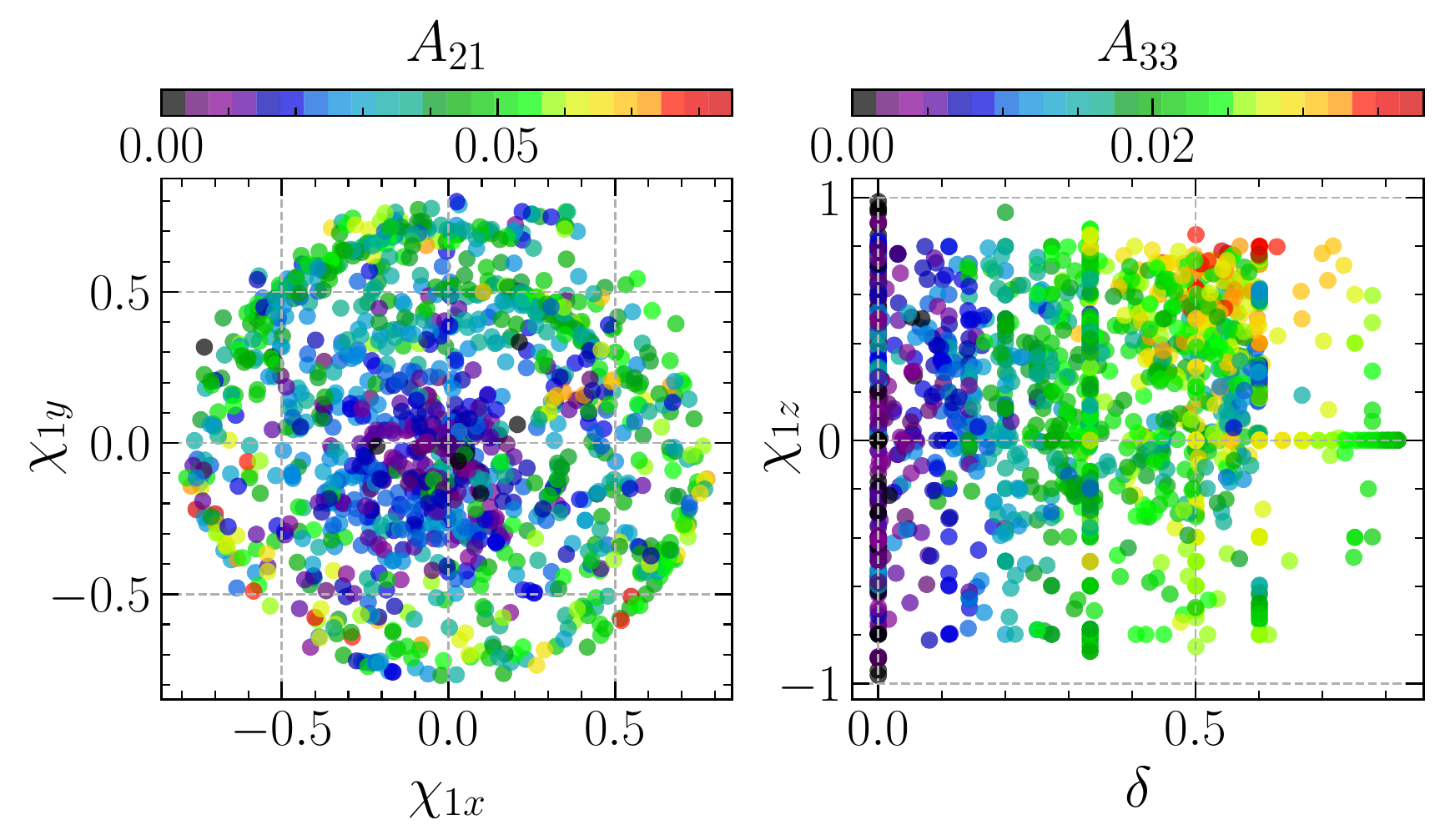}
    \caption{Most significant trends of QNM mode amplitudes in the 7D parameter space made of component spins and asymmetric mass ratio, for for $(2, 1)$ and $(3, 3)$ mode. %
    }
    \label{fig:7d_amp_scatter}
\end{figure}

\subsection{Mode asymmetry}
\label{sec:mode_asymmetry}
In spin-aligned systems, the $m>0$ and $m<0$ modes for a given $l$ are equally excited. However, in precessing systems, this symmetry is broken, in a way that strongly correlates with the remnant kick~\cite{Leong:2025raf, Borchers:2022pah, Mielke:2024kya}. In Fig.~\ref{fig:equat_symm_scatterplots}, we illustrate this point in the context of ringdown amplitudes, focusing on the $(2,\pm2)$, $(2,\pm1)$, and $(3,\pm3)$ modes. When $\phik \simeq \pi$, all modes with $m = l$ exhibit significantly higher excitation in the $m>0$ mode compared to the $m<0$ mode. In contrast, for modes where $m \neq l$, this trend is reversed. Typically, the difference in excitation between $m>0$ and $m<0$ modes is about an order of magnitude smaller than the overall mode amplitudes.

\subsection{7-dimensional parameter space}

Most of the mode amplitudes we analyzed do not exhibit significant variations across the 7D parameter space defined by $\delta$, $\vec{\chi_{1}}$, and $\vec{\chi_{2}}$. However, two notable correlations emerge when examining specific mode amplitudes against these parameters. In line with the phenomenology discussed earlier, we note that high in-plane spins are associated with larger opening and kick angles.

First, the amplitude of the $(2,\pm 1)$ modes strongly depends on the in-plane spin of the larger BH, increasing as the in-plane spin grows. This is illustrated in the left panel of Fig.~\ref{fig:7d_amp_scatter}, where a significant rise in amplitude is observed with increasing radius. Another clear correlation involves the amplitude of the $(3,\pm 3)$ modes, which primarily increases with $\delta$. Additionally, at high values of $\delta$, the amplitude exhibits a notable increase with the $z$-component of the spin of the larger BH.

\section{Fits}
\label{sec:mode_amp_fits}

We interpolate the ringdown amplitudes extracted from NR simulations using GPR~\cite{rasmussen}. GPR allows for smooth, continuous mappings across complex, high-dimensional parameter spaces without relying on an analytical ansatz while also providing uncertainty estimates for the predicted values. Specifically, we present GPR fits to predict nine quantities: the amplitudes $A_{2\pm2}$, $A_{2\pm1}$, $A_{20}^{\rm Re}$, $A_{20}^{\rm Im}$, $A_{3\pm3}$ expressed in the ringdown frame, and $\Delta t_{\rm EMOP}$ (see Appendix~\ref{app:t_emop} for details on the latter). For each quantity, two models are trained using either the 6D or 7D parameter spaces described in Sec.~\ref{sec:parameter_space}, resulting in a total of 18 models.
For a conservative estimate of the uncertainty in our predictions, we train additional GPR models on the Leave-One-Out (LOO) absolute error. Further details about the treatment of uncertainty on GPR prediction are discussed in Appendix~\ref{app:sigma_gpr}. We use the homoscedastic GPR implementation of \textsc{scikit-learn}~\cite{Pedregosa:2011ork}.

\subsection{GPR setup}
\label{sec:GPR_setup}

Gaussian processes are distributions over functions, based on the assumption that any finite subset of points $X = {{\bm x}_1, {\bm x}_2, ..., {\bm x}_n}$ in the data domain is characterized by a multivariate normal joint probability distribution, $f(X) \sim \mathcal{N} \left[ \mu(X), k(X,X) \right]$, where the covariance matrix is defined by the selected kernel function. We use a commonly adopted combination of a radial basis function kernel and a white noise kernel. The former computes the covariance between two points as a function of their Euclidean distance, assigning higher similarity to closer points, while the latter adds a constant variance to the diagonal of the covariance matrix. Our kernel function is
\begin{align}
\label{eq:kernel}
    k(\boldsymbol{x}_i, \boldsymbol{x}_j | a,\boldsymbol{\ell}, N_0) &= a^{2} \prod_{d=1}^D\exp\left[- \frac{(x_{i,d}-x_{j,d})^2}{2\ell_d^2} \right]  
  \notag \\
   &+ (N_{0}+\alpha_{lm}) \; \delta^i_j,
\end{align}
where $D$ is the number of input features (i.e., either 6 or 7 binary BH parameters, depending on the specific case considered), and $\delta^i_j$ denotes the Kronecker delta. 

It is important to remember that the ringdown amplitudes we are fitting here are, in turn, the results of fits to NR simulations, the quality of which varies across the dataset (see Sec.~\ref{sec:amp_extraction}). To address this, we introduce an additional threshold on the waveform fit error for each mode and only consider systems with $\epsNR < 0.08$. Note that this results in models for different modes being trained on slightly different datasets; the number of excluded systems ranges from 0 to 190 depending on the mode. We also incorporate a Tikhonov regularization term into the covariance matrix by adding $\alpha_{lm} = -1/\log{\epsNR}$ to the diagonal. %

As is standard practice in machine learning, we pre-process the input data by normalizing it to have zero mean and unit variance. The training process for GPR involves optimizing the kernel parameters $a$, $\boldsymbol{\ell}$, and $N_0$.

We assess the performance of our GPR model using a LOO cross-validation scheme~\cite{hastie01statisticallearning}. This method is particularly suited to our scenario, given the limitations of our sparse dataset. Traditional data partitioning into test, training, and validation sets would yield an unreliable performance estimate, as results would heavily depend on how sample points are assigned to each subset. %
In the following, LOO is used solely to quantify errors, the fit expectation value is obtained with GPR trained on the full dataset.

For our dataset of size $N$, we perform $N$ training iterations, each time leaving out one data point. For each excluded data point, we compute the corresponding GPR prediction, $A_{lm}^{\rm GPR}$. We evaluate the performance of our GPR model using three metrics: the absolute error, the relative error, and $\epsGPR$, which are defined as follows:
\begin{align}
\mathrm{Err}_{\rm abs} &= |A_{lm} - A_{lm}^{\rm GPR}|\,,\label{eq:abs_err}
\\
\mathrm{Err}_{\rm rel} &= \frac{|A_{lm} - A_{lm}^{\rm GPR}|}{A_{lm}}\,, \label{eq:rel_err}
\\
\epsGPR &= \frac{\int_{t_0}^{100M}\left| h_{lm}^{\rm NR} - h_{lm}^{\rm GPR} \right|^2 dt}{\int_{t_0}^{100M}\left| h_{lm}^{\rm NR} \right|^2 dt}\,.
\end{align}
Here, $h_{lm}^{\rm GPR}$ is the damped sinusoid defined in Eq.\eqref{eq:waveform}, where the amplitude $A_{lm}$ is predicted with GPR, and the phase $\phi_{lm}$ is fixed to the value extracted from the NR waveforms.%
 The $\rm{Err}_{\rm abs}$ and $\rm{Err}_{\rm rel}$ metrics are useful for evaluating the model’s performance in estimating amplitude values, while $\epsGPR$ measures the impact of amplitude uncertainty in the context of a QNM damped sinusoid. This can be compared to $\epsNR$, as defined in Eq.\eqref{eq:eps_nr}.

\begin{figure}
    \centering
    \includegraphics[width=\columnwidth]{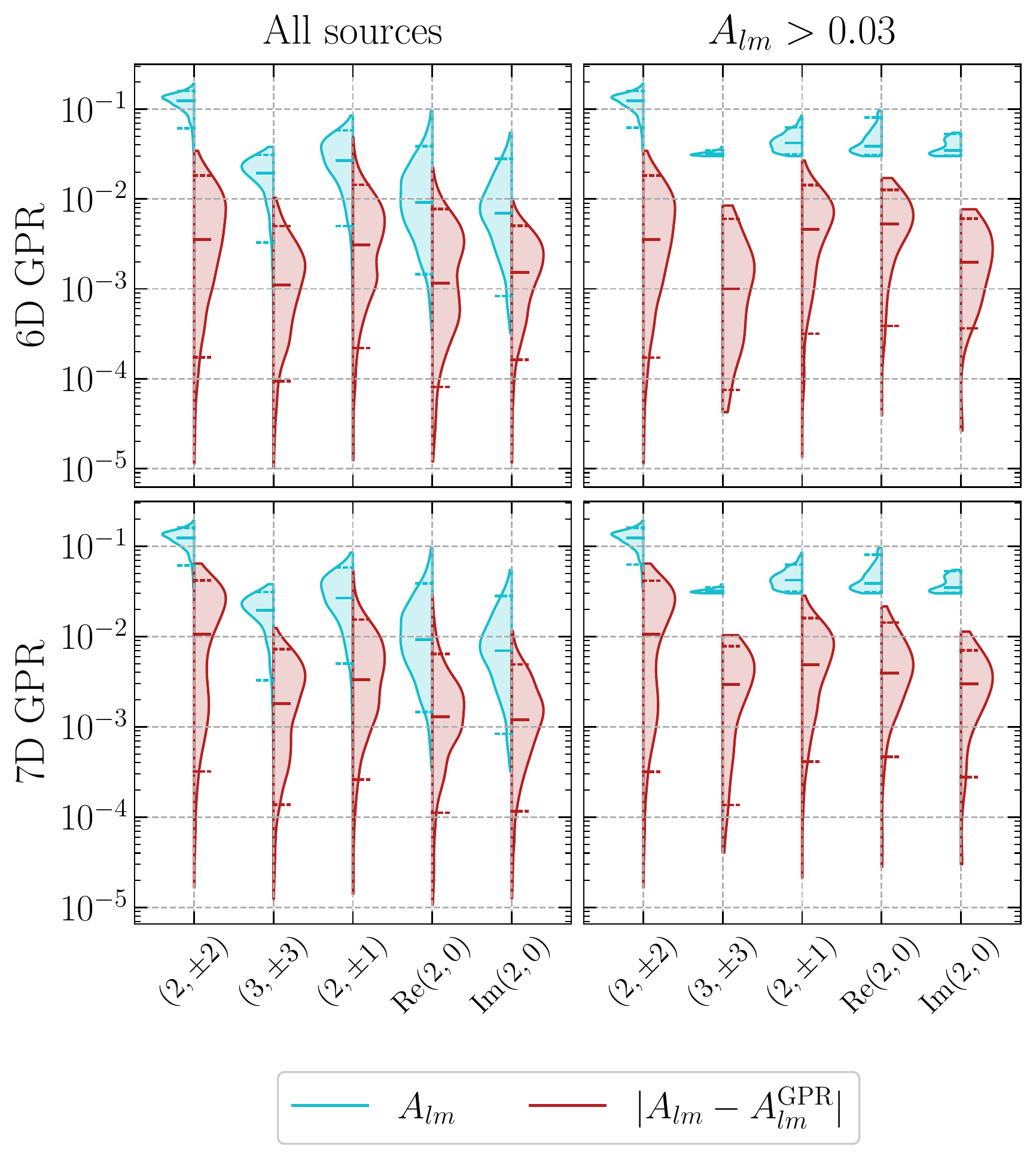}
    \caption{We compare the distributions of the amplitude values extracted from waveforms $A_{lm}$ and the LOO absolute error $|A_{lm} - A_{lm}^{\rm GPR}|$ from the 6D (top) and 7D (bottom) GPR models. These are also reported in Table~\ref{tab:LOO_err}. Solid ticks indicate the median values, while dotted ticks mark the $5\%$ and $95\%$ percentiles. The left panels show data from all sources, while the right panels focus on cases with significantly excited modes, $A_{lm} > 0.03$. %
}
    \label{fig:violin_amp_vs_abs_err_4panel}
\end{figure}

\begin{table*}
\centering
\renewcommand{\arraystretch}{1.5}
\resizebox{\textwidth}{!}{ %
\begin{tabular}{c|c||cc|cc|cc|cc}
\multicolumn{2}{c||}{ } & \multicolumn{2}{c|}{All systems} & \multicolumn{2}{c|}{$A_{lm}>0.03$} & \multicolumn{2}{c|}{Precessing} & \multicolumn{2}{c}{Spin-aligned}\\
 \multicolumn{2}{c||}{ }  & 7D & 6D & 7D & 6D & 7D & 6D & 7D & 6D \\
 \hline \hline
& Abs. &$0.011^{+0.031}_{-0.010}$ & $0.0036^{+0.0147}_{-0.0034}$ & $0.011^{+0.031}_{-0.010}$ & $0.0036^{+0.0146}_{-0.0034}$ & $0.017^{+0.026}_{-0.016}$ & $0.0056^{+0.0147}_{-0.0052}$ & $0.0011^{+0.0025}_{-0.0009}$ & $0.0008^{+0.0027}_{-0.0007}$ \\
$A_{22}$& Rel. &$0.086^{+0.429}_{-0.084}$ & $0.031^{+0.155}_{-0.030}$ & $0.085^{+0.414}_{-0.083}$ & $0.031^{+0.154}_{-0.030}$ & $0.14^{+0.46}_{-0.14}$ & $0.049^{+0.164}_{-0.046}$ & $0.0088^{+0.0301}_{-0.0077}$ & $0.0067^{+0.0279}_{-0.0062}$ \\
& $\epsGPR$ &$0.011^{+0.233}_{-0.008}$ & $0.0049^{+0.0316}_{-0.0024}$ & $0.011^{+0.221}_{-0.008}$ & $0.0049^{+0.0307}_{-0.0024}$ & $0.024^{+0.321}_{-0.021}$ & $0.0061^{+0.0419}_{-0.0032}$ & $0.0034^{+0.0027}_{-0.0012}$ & $0.0033^{+0.0030}_{-0.0013}$ \\
& $\epsNR$ &$0.0007^{+0.0014}_{-0.0004}$ & $0.0007^{+0.0014}_{-0.0004}$ & $0.0007^{+0.0014}_{-0.0004}$ & $0.0007^{+0.0014}_{-0.0004}$ & $0.0007^{+0.0013}_{-0.0005}$ & $0.0007^{+0.0013}_{-0.0005}$ & $0.0006^{+0.0013}_{-0.0002}$ & $0.0006^{+0.0013}_{-0.0002}$ \\
\hline
& Abs. &$0.0099^{+0.0314}_{-0.0095}$ & $0.0034^{+0.0154}_{-0.0033}$ & $0.0099^{+0.0314}_{-0.0095}$ & $0.0034^{+0.0154}_{-0.0033}$ & $0.018^{+0.026}_{-0.017}$ & $0.0056^{+0.0148}_{-0.0052}$ & $0.0014^{+0.0033}_{-0.0012}$ & $0.0009^{+0.0030}_{-0.0008}$ \\
$A_{2-2}$& Rel. &$0.089^{+0.408}_{-0.086}$ & $0.032^{+0.167}_{-0.031}$ & $0.088^{+0.405}_{-0.085}$ & $0.032^{+0.164}_{-0.030}$ & $0.15^{+0.44}_{-0.14}$ & $0.050^{+0.168}_{-0.046}$ & $0.011^{+0.038}_{-0.010}$ & $0.0074^{+0.0313}_{-0.0069}$ \\
& $\epsGPR$ &$0.011^{+0.213}_{-0.009}$ & $0.0049^{+0.0360}_{-0.0025}$ & $0.011^{+0.210}_{-0.008}$ & $0.0049^{+0.0344}_{-0.0025}$ & $0.026^{+0.306}_{-0.023}$ & $0.0062^{+0.0424}_{-0.0034}$ & $0.0035^{+0.0036}_{-0.0016}$ & $0.0033^{+0.0029}_{-0.0015}$ \\
& $\epsNR$ &$0.0006^{+0.0015}_{-0.0004}$ & $0.0006^{+0.0015}_{-0.0004}$ & $0.0006^{+0.0015}_{-0.0004}$ & $0.0006^{+0.0015}_{-0.0004}$ & $0.0007^{+0.0016}_{-0.0005}$ & $0.0007^{+0.0016}_{-0.0005}$ & $0.0006^{+0.0013}_{-0.0002}$ & $0.0006^{+0.0013}_{-0.0002}$ \\
\hline
& Abs. &$0.0019^{+0.0053}_{-0.0017}$ & $0.0011^{+0.0038}_{-0.0010}$ & $0.0029^{+0.0049}_{-0.0028}$ & $0.001^{+0.005}_{-0.001}$ & $0.0026^{+0.0050}_{-0.0024}$ & $0.0015^{+0.0043}_{-0.0013}$ & $0.0005^{+0.0024}_{-0.0005}$ & $0.0004^{+0.0019}_{-0.0004}$ \\
$A_{33}$& Rel. &$0.11^{+0.60}_{-0.11}$ & $0.064^{+0.484}_{-0.060}$ & $0.088^{+0.146}_{-0.083}$ & $0.031^{+0.162}_{-0.029}$ & $0.15^{+0.59}_{-0.14}$ & $0.086^{+0.420}_{-0.079}$ & $0.022^{+0.613}_{-0.019}$ & $0.021^{+0.577}_{-0.019}$ \\
& $\epsGPR$ &$0.024^{+1.395}_{-0.018}$ & $0.014^{+0.599}_{-0.008}$ & $0.020^{+0.045}_{-0.012}$ & $0.012^{+0.034}_{-0.005}$ & $0.031^{+0.499}_{-0.024}$ & $0.016^{+0.262}_{-0.010}$ & $0.0096^{+1.1e+08}_{-0.0046}$ & $0.0093^{+1.7e+07}_{-0.0041}$ \\
& $\epsNR$ &$0.0015^{+0.0132}_{-0.0011}$ & $0.0015^{+0.0132}_{-0.0011}$ & $0.0021^{+0.0027}_{-0.0010}$ & $0.0021^{+0.0027}_{-0.0010}$ & $0.0016^{+0.0114}_{-0.0011}$ & $0.0016^{+0.0114}_{-0.0011}$ & $0.0014^{+0.0157}_{-0.0011}$ & $0.0014^{+0.0157}_{-0.0011}$ \\
\hline
& Abs. &$0.0017^{+0.0053}_{-0.0016}$ & $0.0011^{+0.0037}_{-0.0011}$ & $0.0026^{+0.0057}_{-0.0025}$ & $0.0011^{+0.0040}_{-0.0010}$ & $0.0023^{+0.0050}_{-0.0021}$ & $0.0015^{+0.0038}_{-0.0014}$ & $0.0007^{+0.0022}_{-0.0006}$ & $0.0005^{+0.0020}_{-0.0005}$ \\
$A_{3-3}$& Rel. &$0.10^{+0.63}_{-0.09}$ & $0.069^{+0.499}_{-0.065}$ & $0.080^{+0.171}_{-0.077}$ & $0.034^{+0.120}_{-0.031}$ & $0.14^{+0.61}_{-0.13}$ & $0.084^{+0.469}_{-0.077}$ & $0.030^{+0.606}_{-0.026}$ & $0.023^{+0.594}_{-0.021}$ \\
& $\epsGPR$ &$0.022^{+1.141}_{-0.015}$ & $0.015^{+0.883}_{-0.009}$ & $0.019^{+0.056}_{-0.010}$ & $0.013^{+0.022}_{-0.005}$ & $0.029^{+0.509}_{-0.021}$ & $0.016^{+0.290}_{-0.009}$ & $0.0099^{+8.6e+07}_{-0.0046}$ & $0.0095^{+3.9e+07}_{-0.0045}$ \\
& $\epsNR$ &$0.0015^{+0.0154}_{-0.0010}$ & $0.0015^{+0.0154}_{-0.0010}$ & $0.0021^{+0.0027}_{-0.0011}$ & $0.0021^{+0.0027}_{-0.0011}$ & $0.0015^{+0.0148}_{-0.0010}$ & $0.0015^{+0.0148}_{-0.0010}$ & $0.0013^{+0.0160}_{-0.0011}$ & $0.0013^{+0.0160}_{-0.0011}$ \\
\hline
& Abs. &$0.0034^{+0.0118}_{-0.0031}$ & $0.0030^{+0.0113}_{-0.0028}$ & $0.0048^{+0.0111}_{-0.0044}$ & $0.0046^{+0.0097}_{-0.0043}$ & $0.0047^{+0.0123}_{-0.0042}$ & $0.0045^{+0.0114}_{-0.0041}$ & $0.0012^{+0.0042}_{-0.0011}$ & $0.0010^{+0.0029}_{-0.0009}$ \\
$A_{21}$& Rel. &$0.14^{+0.98}_{-0.13}$ & $0.13^{+0.99}_{-0.12}$ & $0.11^{+0.23}_{-0.10}$ & $0.11^{+0.18}_{-0.10}$ & $0.17^{+0.84}_{-0.15}$ & $0.15^{+0.97}_{-0.14}$ & $0.058^{+1.380}_{-0.052}$ & $0.060^{+1.006}_{-0.055}$ \\
& $\epsGPR$ &$0.027^{+3.814}_{-0.024}$ & $0.023^{+3.160}_{-0.020}$ & $0.021^{+0.100}_{-0.018}$ & $0.020^{+0.089}_{-0.017}$ & $0.032^{+0.936}_{-0.029}$ & $0.031^{+1.157}_{-0.027}$ & $0.0091^{+5.6e+08}_{-0.0058}$ & $0.0087^{+3.3e+08}_{-0.0054}$ \\
& $\epsNR$ &$0.0015^{+0.0296}_{-0.0011}$ & $0.0015^{+0.0296}_{-0.0011}$ & $0.0018^{+0.0461}_{-0.0014}$ & $0.0018^{+0.0461}_{-0.0014}$ & $0.0014^{+0.0371}_{-0.0010}$ & $0.0014^{+0.0371}_{-0.0010}$ & $0.0016^{+0.0059}_{-0.0015}$ & $0.0016^{+0.0059}_{-0.0015}$ \\
\hline
& Abs. &$0.0034^{+0.0117}_{-0.0032}$ & $0.0026^{+0.0104}_{-0.0024}$ & $0.0051^{+0.0111}_{-0.0048}$ & $0.0041^{+0.0096}_{-0.0038}$ & $0.0046^{+0.0119}_{-0.0042}$ & $0.0040^{+0.0103}_{-0.0037}$ & $0.0012^{+0.0040}_{-0.0011}$ & $0.0007^{+0.0020}_{-0.0006}$ \\
$A_{2-1}$& Rel. &$0.14^{+1.03}_{-0.13}$ & $0.11^{+0.75}_{-0.10}$ & $0.12^{+0.24}_{-0.11}$ & $0.096^{+0.177}_{-0.090}$ & $0.16^{+0.84}_{-0.15}$ & $0.14^{+0.71}_{-0.12}$ & $0.056^{+1.619}_{-0.053}$ & $0.040^{+0.879}_{-0.036}$ \\
& $\epsGPR$ &$0.026^{+4.044}_{-0.022}$ & $0.018^{+2.569}_{-0.015}$ & $0.021^{+0.111}_{-0.018}$ & $0.016^{+0.088}_{-0.013}$ & $0.031^{+0.890}_{-0.028}$ & $0.026^{+0.632}_{-0.023}$ & $0.0086^{+5.9e+08}_{-0.0054}$ & $0.0064^{+4.8e+07}_{-0.0033}$ \\
& $\epsNR$ &$0.0014^{+0.0361}_{-0.0011}$ & $0.0014^{+0.0361}_{-0.0011}$ & $0.0017^{+0.0603}_{-0.0013}$ & $0.0017^{+0.0603}_{-0.0013}$ & $0.0012^{+0.0465}_{-0.0009}$ & $0.0012^{+0.0465}_{-0.0009}$ & $0.0016^{+0.0059}_{-0.0015}$ & $0.0016^{+0.0059}_{-0.0015}$ \\
\hline
& Abs. &$0.0013^{+0.0050}_{-0.0012}$ & $0.0011^{+0.0066}_{-0.0011}$ & $0.0039^{+0.0103}_{-0.0035}$ & $0.0052^{+0.0074}_{-0.0049}$ & $0.0015^{+0.0055}_{-0.0014}$ & $0.0021^{+0.0062}_{-0.0020}$ & $0.0009^{+0.0020}_{-0.0008}$ & $0.0004^{+0.0009}_{-0.0004}$ \\
$A_{20}^{\rm Re}$& Rel. &$0.13^{+1.08}_{-0.12}$ & $0.13^{+1.10}_{-0.12}$ & $0.10^{+0.13}_{-0.09}$ & $0.12^{+0.19}_{-0.11}$ & $0.13^{+1.06}_{-0.12}$ & $0.17^{+1.30}_{-0.16}$ & $0.11^{+1.13}_{-0.11}$ & $0.062^{+0.277}_{-0.056}$ \\
& $\epsGPR$ &$0.026^{+1.625}_{-0.023}$ & $0.032^{+1.331}_{-0.029}$ & $0.012^{+0.045}_{-0.010}$ & $0.016^{+0.080}_{-0.014}$ & $0.026^{+1.358}_{-0.023}$ & $0.045^{+1.818}_{-0.042}$ & $0.026^{+1.960}_{-0.019}$ & $0.016^{+0.976}_{-0.013}$ \\
& $\epsNR$ &$0.0043^{+0.2504}_{-0.0042}$ & $0.0043^{+0.2504}_{-0.0042}$ & $0.0003^{+0.0066}_{-0.0002}$ & $0.0003^{+0.0066}_{-0.0002}$ & $0.0027^{+0.2042}_{-0.0025}$ & $0.0027^{+0.2042}_{-0.0025}$ & $0.0073^{+0.3650}_{-0.0062}$ & $0.0073^{+0.3650}_{-0.0062}$ \\
\hline
& Abs. &$0.0008^{+0.0036}_{-0.0007}$ & $0.0008^{+0.0037}_{-0.0008}$ & $0.0030^{+0.0040}_{-0.0027}$ & $0.0020^{+0.0041}_{-0.0016}$ & $0.0011^{+0.0037}_{-0.0010}$ & $0.0015^{+0.0035}_{-0.0014}$ & $0.0002^{+0.0011}_{-0.0002}$ & $0.0002^{+0.0003}_{-0.0001}$ \\
$A_{20}^{\rm Im}$& Rel. &$0.19^{+0.83}_{-0.17}$ & $0.21^{+1.89}_{-0.20}$ & $0.074^{+0.140}_{-0.065}$ & $0.052^{+0.119}_{-0.043}$ & $0.19^{+0.83}_{-0.17}$ & $0.21^{+1.89}_{-0.20}$ & - & - \\
& $\epsGPR$ &$0.13^{+8e+09}_{-0.12}$ & $0.16^{+3.4e+09}_{-0.15}$ & $0.0079^{+0.0329}_{-0.0061}$ & $0.0049^{+0.0301}_{-0.0031}$ & $0.055^{+1.740}_{-0.051}$ & $0.066^{+9.217}_{-0.061}$ & - & - \\
& $\epsNR$ &$0.0021^{+0.1121}_{-0.0021}$ & $0.0021^{+0.1121}_{-0.0021}$ & $0.0006^{+0.0016}_{-0.0004}$ & $0.0006^{+0.0016}_{-0.0004}$ & $0.0048^{+0.1351}_{-0.0045}$ & $0.0048^{+0.1351}_{-0.0045}$ & - & - \\

\end{tabular}
} %
\renewcommand{\arraystretch}{1}
\caption{Performance metrics of our GPR fits. We list median values, along with their distance from 5\textsuperscript{th} and 95\textsuperscript{th} percentiles for amplitude absolute error (Abs.), relative error (Rel.), $\epsGPR$, and $\epsNR$. Some values are missing because for spin-aligned systems all points have $A_{20}^{\rm Im} = 0$. We exclude simulations with $A_{lm} = 0$ when evaluating relative errors.}
\label{tab:LOO_err}
\end{table*}

\subsection{Model performance}
\label{sec:gpr_performance_singlemodes}

In Table~\ref{tab:LOO_err}, we summarize the performance of our models using the LOO framework. For each mode, we report the median, $5^{\rm th}$, and $95^{\rm th}$ percentiles of the absolute error, relative error, and the error statistics $\epsGPR$ and $\epsNR$. These performance metrics are computed for the entire dataset as well as specific subsets. For significantly excited modes ($A_{lm} > 0.03$), accurate predictions are particularly important, as less excited modes are inherently more challenging to detect.

In Fig.~\ref{fig:violin_amp_vs_abs_err_4panel}, we show a comparison between distributions of amplitude values extracted from waveforms $A_{lm}$, and LOO absolute error from Eq.\eqref{eq:abs_err} for the 6D and 7D GPR models. %
The performance of our 6D and 7D models is comparable between $m>0$ and $m<0$ modes. The 6D models perform better for the $(2, \pm 2)$  modes across the entire dataset, achieving a median relative error of 3\%. This is followed by the $(3, \pm 3)$ modes at 7\%, and finally $(2, \pm 1)$ and $\rm{Re}(2, 0)$ modes at $\sim 13\%$. For all modes, the subset with $A_{lm} > 0.03$ exhibits lower amplitude relative error and $\epsGPR$, indicating improved performance for significantly excited modes. %
For instance, the median relative error improves from $7\%$ to $3\%$ for $(3, \pm 3)$ and  from $11\%-13\%$ to $10\%-11\%$ for $(2, \pm 1)$.

The performance of the 6D models is generally better than or comparable to that of the 7D models, reflecting the advantage of selecting parameters along which the amplitudes vary more smoothly. This performance difference is particularly evident in the precessing subset for the $(2, \pm 2)$ mode, where the median relative error for the 7D model is approximately three times worse than that of the 6D model (5\% to 15\%), and in the $(3, \pm 3)$ modes, where the error is roughly twice as large (8\% to 15\%). Interestingly, for $(2, \pm 1)$ and $\rm{Re}(2, 0)$ modes—where the amplitudes vary smoothly with the in-plane spin components $\chi_{1x}$ and $\chi_{1y}$—both the 6D and 7D models perform comparably. This is also the case for the spin-aligned subset of sources, where the ringdown amplitudes vary smoothly with the $z$-component of the spins. 

Our findings highlight the importance of selecting appropriate parameter combinations that optimize prediction accuracy. Exploring such optimal parametrizations is an important avenue for future work and could lead to significant improvements in model performance.

\subsection{Comparison with other models}
\label{sec:comparison_previous_models}

\begin{figure}
    \centering
    \includegraphics[width=\columnwidth]{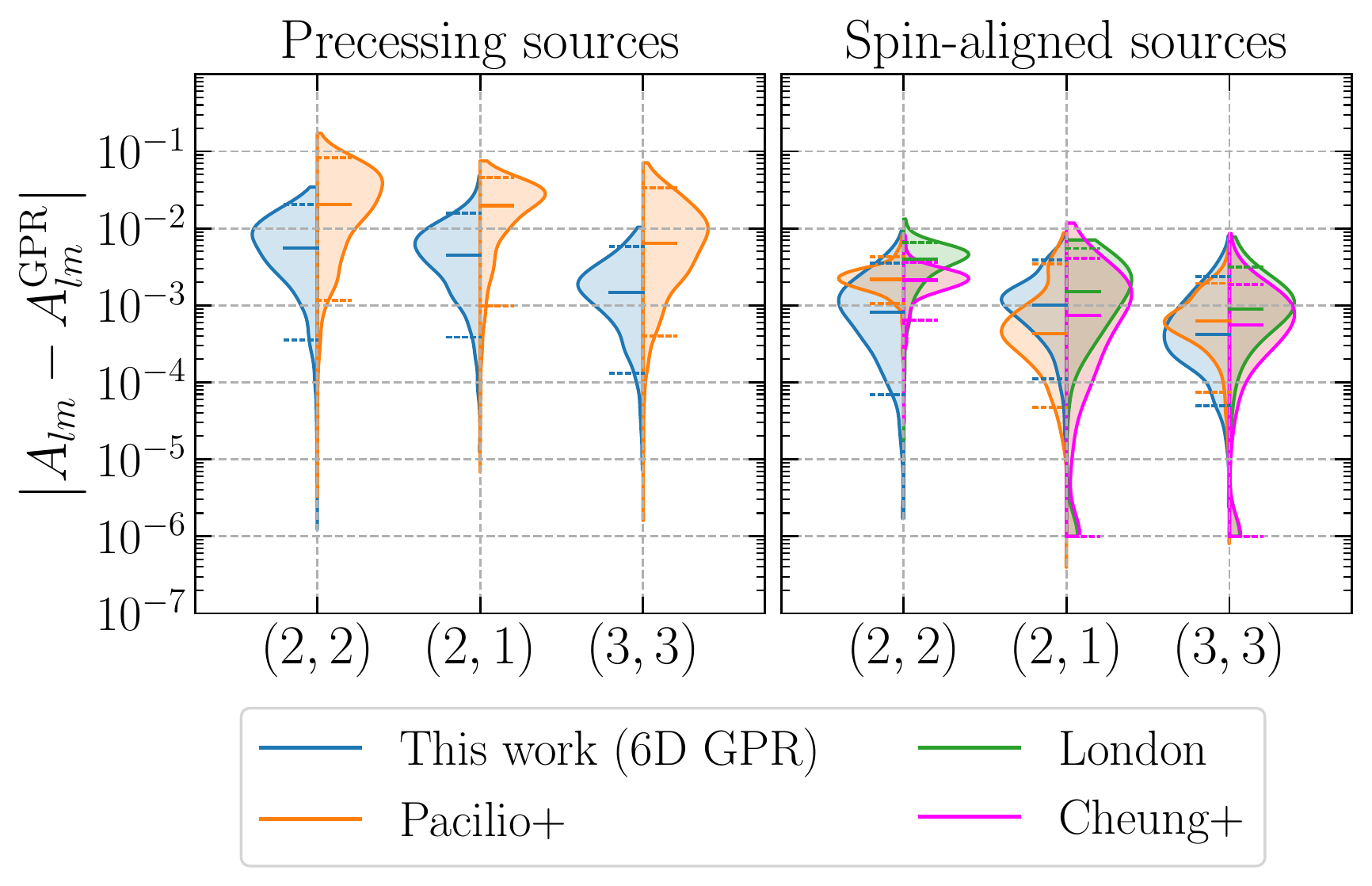}
    \caption{Comparison of absolute error distributions for precessing (left) and spin-aligned (right) sources, using our 6D GPR model and other models presented in recent literature. In the legend, ``Pacilio+'' refers to Ref.~\cite{Pacilio:2024tdl}, ``London'' refers to Ref.~\cite{London:2018gaq}, and ``Cheung+'' refers to Ref.~\cite{Cheung:2023vki}.
    }
    \label{fig:compare_models_violin}
\end{figure}

\begin{table}[h]
\centering
\renewcommand{\arraystretch}{2.5}

\resizebox{\columnwidth}{!}{
\begin{tabular}{c|c||c|c|c|c}
\multicolumn{2}{c||}{ } & \multicolumn{2}{c|}{Absolute error} & \multicolumn{2}{c}{Relative error} \\
\cline{3-6}
\multicolumn{2}{c||}{ } & Precessing fit& Aligned fit & Precessing fit& Aligned fit\\
\hline
\hline

\multirow{3}{*}{\rotatebox{90}{Precessing sources}} 
& $A_{22}$  & $0.0056^{+0.0147}_{-0.0052}$ & $0.020^{+0.062}_{-0.019}$ & $0.049^{+0.164}_{-0.046}$ & $0.18^{+0.64}_{-0.17}$ \\
& $A_{21}$  & $0.0045^{+0.0114}_{-0.0041}$ & $0.020^{+0.026}_{-0.019}$ & $0.15^{+0.97}_{-0.14}$ & $0.69^{+0.28}_{-0.62}$ \\
& $A_{33}$  & $0.0015^{+0.0045}_{-0.0013}$ & $0.0063^{+0.0273}_{-0.0059}$ & $0.086^{+0.420}_{-0.079}$ & $0.47^{+1.50}_{-0.45}$ \\
\cline{1-6}
\multirow{3}{*}{\rotatebox{90}{Aligned sources}}  
& $A_{22}$  & $0.0008^{+0.0027}_{-0.0007}$ & $0.0022^{+0.0021}_{-0.0011}$ & $0.0067^{+0.0279}_{-0.0062}$ & $0.018^{+0.017}_{-0.008}$ \\
& $A_{21}$  & $0.0010^{+0.0029}_{-0.0009}$ & $0.0004^{+0.0030}_{-0.0004}$ & $0.060^{+1.006}_{-0.055}$ & $0.023^{+0.685}_{-0.020}$ \\
& $A_{33}$  & $0.0004^{+0.0019}_{-0.0004}$ & $0.0006^{+0.0013}_{-0.0005}$ & $0.021^{+0.577}_{-0.019}$ & $0.030^{+0.285}_{-0.023}$ \\
\end{tabular}
}
\caption{%
Comparison between our previous ringdown fit from Ref.~\cite{Pacilio:2024tdl}, which was calibrated on spin aligned sources, and the precessing 6D fit presented in this paper. 
The columns in this table indicate which GPR model is used (aligned or precessing), while the rows specify the target source for amplitude prediction. We present both absolute and relative errors and list median values along with the distances to the $5^{\rm th}$ and $95^{\rm th}$ percentiles.}
\label{tab:comparison_aligned_vs_prec}
\end{table}

In Fig.~\ref{fig:compare_models_violin}, we compare the results of our 6D GPR models with previous results from both our own work (``Pacilio+''~\cite{Pacilio:2024tdl}) and other authors (``Cheung+''~\cite{Cheung:2023vki}, ``London''~\cite{London:2018gaq}). The former is a GPR model while the latter are polynomial fits. These were all constructed for spin-aligned systems and did not attempt to capture precession effects.

When comparing with our previous work~\cite{Pacilio:2024tdl}, we use both models to predict amplitudes for both the precessing and spin-aligned subsets of data, even though the former model was only calibrated on spin-aligned systems. 
Considering precessing sources, we observe that our 6D model trained on generic spins yields a substantial improvement in amplitude prediction accuracy for all modes. Specifically, the median relative error for $A_{22}$ reduces from approximately 18\% to 5\%, for $A_{21}$ from 70\% to 15\%, and for $A_{33}$ from 47\% to 9\%.

More details on this comparison are provided in Table~\ref{tab:comparison_aligned_vs_prec}.
Specifically, we test our aligned spin fit from Ref.~\cite{Pacilio:2024tdl} outside its intended scope to assess whether our 6D GPR model offers greater predictive accuracy. This is particularly relevant because a common simplifying assumption is to ignore non-aligned spin components and treat ringdowns as if they were generated from mergers of  binaries with aligned spins. Table~\ref{tab:comparison_aligned_vs_prec} quantifies the potential improvement in prediction accuracy that can be obtained when using our precessing 6D GPR model instead of this conventional assumption.

Restricting to spin-aligned systems, Fig.~\ref{fig:compare_models_violin} compares the absolute error distributions of our 6D model against those of Refs.~\cite{Pacilio:2024tdl,Cheung:2022rbm,London:2018gaq}. The absolute error values for all models are one or two orders of magnitude smaller than the related amplitude value range. The differences between the models are not significant, and most likely stem from different waveform fitting techniques employed to extract amplitude values from NR simulations. From this comparison, we conclude that GPR models are robust when applied to spin-aligned systems, with performances similar to current state-of-the-art methods. Crucially, we now capture precessing spins.

\begin{figure}
    \centering
    \includegraphics[width=\columnwidth]{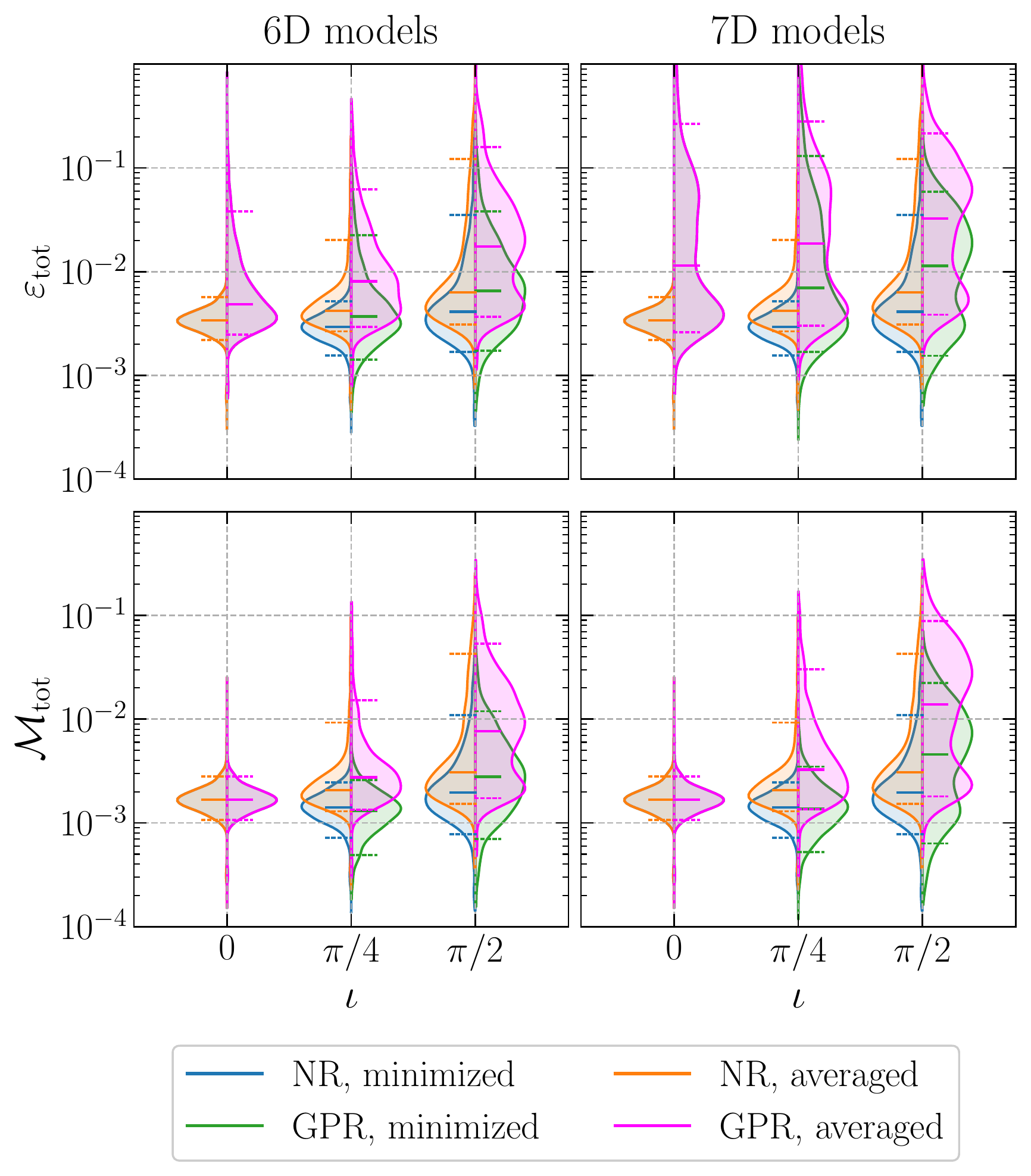}
    \caption{Distributions of the strain relative errors (top panels) and mismatches (bottom panels), computed for both the GPR fit and the waveform extracted from NR simulations, are presented. We report results where we either average or minimize over the angle $\varphi$ entering the spherical harmonics. Solid ticks indicate the median values, while dotted ticks mark the $5\%$ and $95\%$ percentiles. The left and right panels correspond to GPR fits in our 6D and 7D parameter spaces, respectively. Complementary results are reported in Table~\ref{tab:tot_mismatch}. 
}
\label{fig:violin_tot_waveform_eps_mism_4panel}
\end{figure}

\begin{table*}
\centering
\renewcommand{\arraystretch}{1.5}
\begin{tabular}{c|c||cc|cc|cc}
\multicolumn{1}{c|}{ } & \multicolumn{1}{c||}{ } & \multicolumn{2}{c|}{All systems} & \multicolumn{2}{c|}{Precessing} & \multicolumn{2}{c}{Spin-aligned}\\
 \multicolumn{1}{c|}{ } & $\iota$  & 7D & 6D & 7D & 6D & 7D & 6D \\
 \hline \hline

& $0$ &$0.011^{+0.254}_{-0.009}$ & $0.0049^{+0.0333}_{-0.0024}$ & $0.026^{+0.339}_{-0.022}$ & $0.0063^{+0.0445}_{-0.0034}$ & $0.0034^{+0.0027}_{-0.0012}$ & $0.0033^{+0.0031}_{-0.0013}$ \\
$\langle\epsGPRtot(\iota)\rangle_\varphi$& $\pi/4$ &$0.019^{+0.260}_{-0.016}$ & $0.0081^{+0.0543}_{-0.0051}$ & $0.035^{+0.318}_{-0.030}$ & $0.011^{+0.069}_{-0.007}$ & $0.0039^{+0.0034}_{-0.0012}$ & $0.0037^{+0.0037}_{-0.0011}$ \\
& $\pi/2$ &$0.032^{+0.182}_{-0.029}$ & $0.017^{+0.141}_{-0.014}$ & $0.055^{+0.189}_{-0.048}$ & $0.026^{+0.178}_{-0.020}$ & $0.0055^{+0.0071}_{-0.0021}$ & $0.0046^{+0.0053}_{-0.0013}$ \\
\hline
& $0$ &$0.0034^{+0.0023}_{-0.0012}$ & $0.0034^{+0.0023}_{-0.0012}$ & $0.0035^{+0.0022}_{-0.0011}$ & $0.0035^{+0.0022}_{-0.0011}$ & $0.0032^{+0.0025}_{-0.0012}$ & $0.0032^{+0.0025}_{-0.0012}$ \\
$\langle\epsNRtot(\iota)\rangle_\varphi$& $\pi/4$ &$0.0042^{+0.0160}_{-0.0016}$ & $0.0042^{+0.0160}_{-0.0016}$ & $0.0046^{+0.0249}_{-0.0016}$ & $0.0046^{+0.0249}_{-0.0016}$ & $0.0034^{+0.0029}_{-0.0010}$ & $0.0034^{+0.0029}_{-0.0010}$ \\
& $\pi/2$ &$0.0063^{+0.1157}_{-0.0032}$ & $0.0063^{+0.1157}_{-0.0032}$ & $0.0084^{+0.1474}_{-0.0048}$ & $0.0084^{+0.1474}_{-0.0048}$ & $0.0039^{+0.0040}_{-0.0014}$ & $0.0039^{+0.0040}_{-0.0014}$ \\
\hline
& $0$ &$0.0017^{+0.0011}_{-0.0006}$ & $0.0017^{+0.0011}_{-0.0006}$ & $0.0017^{+0.0011}_{-0.0006}$ & $0.0017^{+0.0011}_{-0.0006}$ & $0.0016^{+0.0013}_{-0.0006}$ & $0.0016^{+0.0013}_{-0.0006}$ \\
$\langle\mismGPRtot(\iota)\rangle_\varphi$& $\pi/4$ &$0.0033^{+0.0270}_{-0.0019}$ & $0.0028^{+0.0125}_{-0.0014}$ & $0.0044^{+0.0329}_{-0.0027}$ & $0.0032^{+0.0158}_{-0.0015}$ & $0.0017^{+0.0017}_{-0.0005}$ & $0.0017^{+0.0016}_{-0.0005}$ \\
& $\pi/2$ &$0.014^{+0.074}_{-0.012}$ & $0.0077^{+0.0458}_{-0.0059}$ & $0.024^{+0.078}_{-0.021}$ & $0.011^{+0.057}_{-0.008}$ & $0.0023^{+0.0027}_{-0.0008}$ & $0.0021^{+0.0021}_{-0.0006}$ \\
\hline
& $0$ &$0.0017^{+0.0011}_{-0.0006}$ & $0.0017^{+0.0011}_{-0.0006}$ & $0.0017^{+0.0011}_{-0.0006}$ & $0.0017^{+0.0011}_{-0.0006}$ & $0.0016^{+0.0013}_{-0.0006}$ & $0.0016^{+0.0013}_{-0.0006}$ \\
$\langle\mismNRtot(\iota)\rangle_\varphi$& $\pi/4$ &$0.0021^{+0.0072}_{-0.0008}$ & $0.0021^{+0.0072}_{-0.0008}$ & $0.0023^{+0.0103}_{-0.0008}$ & $0.0023^{+0.0103}_{-0.0008}$ & $0.0017^{+0.0015}_{-0.0005}$ & $0.0017^{+0.0015}_{-0.0005}$ \\
& $\pi/2$ &$0.0031^{+0.0396}_{-0.0016}$ & $0.0031^{+0.0396}_{-0.0016}$ & $0.0040^{+0.0494}_{-0.0022}$ & $0.0040^{+0.0494}_{-0.0022}$ & $0.0019^{+0.0017}_{-0.0006}$ & $0.0019^{+0.0017}_{-0.0006}$ \\
\hline

\end{tabular}
\renewcommand{\arraystretch}{1}
\caption{%
Performance of our GPR models in terms of the all-mode relative error and the mismatch. We present results for representative inclination angles. Specifically, we report $\langle\epsGPRtot\rangle_\varphi$ and $\langle\mismGPRtot\rangle_\varphi$ to highlight the accuracy of the GPR models. The metrics $\langle\epsNRtot\rangle_\varphi$ and $\langle\mismNRtot\rangle_\varphi$ serve as baselines for interpreting GPR performance.
} 
\label{tab:tot_mismatch}
\end{table*}

\subsection{Mismatches}

We evaluate the impact of our prediction on GW data analysis by computing mismatches. 
The GW strain is given by 
\begin{equation}
    h (t;\iota, \varphi) = \sum_{lm} h_{lm}(t) {}_{-2}Y_{lm}(\iota, \varphi) ,
\label{eq:mode_combination}
\end{equation}
where ${}_{-2}Y_{lm}(\iota, \varphi)$ are the spin-weighted spherical harmonics. For ease of reading, we will drop the dependencies of $h(t; \iota, \varphi)$ from now on.
We use two performance metrics to quantify the difference between the analytical ringdown waveform obtained using our predicted amplitudes and the NR waveform, namely the relative error $\epsGPRtot$ and the mismatch $\mismGPRtot$:
\begin{align} 
\epsGPRtot(\iota,\varphi) &= \frac{\int_{t_0}^{100M}\left| h^{\rm NR} - h^{\rm GPR} \right|^2 dt}{\int_{t_0}^{100M}\left| h^{\rm NR} \right|^2 dt},
\label{eq:epsGPRtot}
\\
\mismGPRtot(\iota,\varphi) &= 1 - \frac{\braket{h^{\rm NR} | h^{\rm GPR}}}{
\sqrt{\braket{h^{\rm NR} | h^{\rm NR}} 
\braket{h^{\rm GPR} | h^{\rm GPR}}}}.
\label{eq:mismatch_gpr}
\end{align}
The inner product is computed in the time domain assuming flat power-spectral density:
\begin{align}
    \label{eq:inner:product}
    \braket{a|b}= {\rm Re}\int a(t)b^*(t)\, \dd t \,.
\end{align}
Here, $h^{\rm NR}$ refers to the NR waveform, while $h^{\rm GPR}$ is the waveform obtained for each system using QNM amplitudes predicted by our GPR models, with QNM phases fixed to values extracted from fitting the NR waveforms as described in Sec.~\ref{sec:amp_extraction}. As in previous analyses (e.g. Ref.~\cite{Cotesta:2018fcv}), we average over the spherical harmonics angle $\varphi$; the resulting metrics are denoted $\langle\mismGPRtot(\iota)\rangle_\varphi$, $\langle\epsGPRtot(\iota)\rangle_\varphi$.
Similarly, we compute $\langle\mismNRtot(\iota)\rangle_\varphi$, $\langle\epsNRtot(\iota)\rangle_\varphi$, using Eqs.~\eqref{eq:epsGPRtot}-\eqref{eq:mismatch_gpr}, replacing $h^{\rm GPR}$ with $h^{\rm fit}$, i.e. the waveform obtained for each system by setting both QNM amplitudes and phases to values extracted from fitting the NR waveforms as described in Sec.~\ref{sec:amp_extraction}.

 \begin{figure*}
    \centering
    \includegraphics[width=1.8\columnwidth]{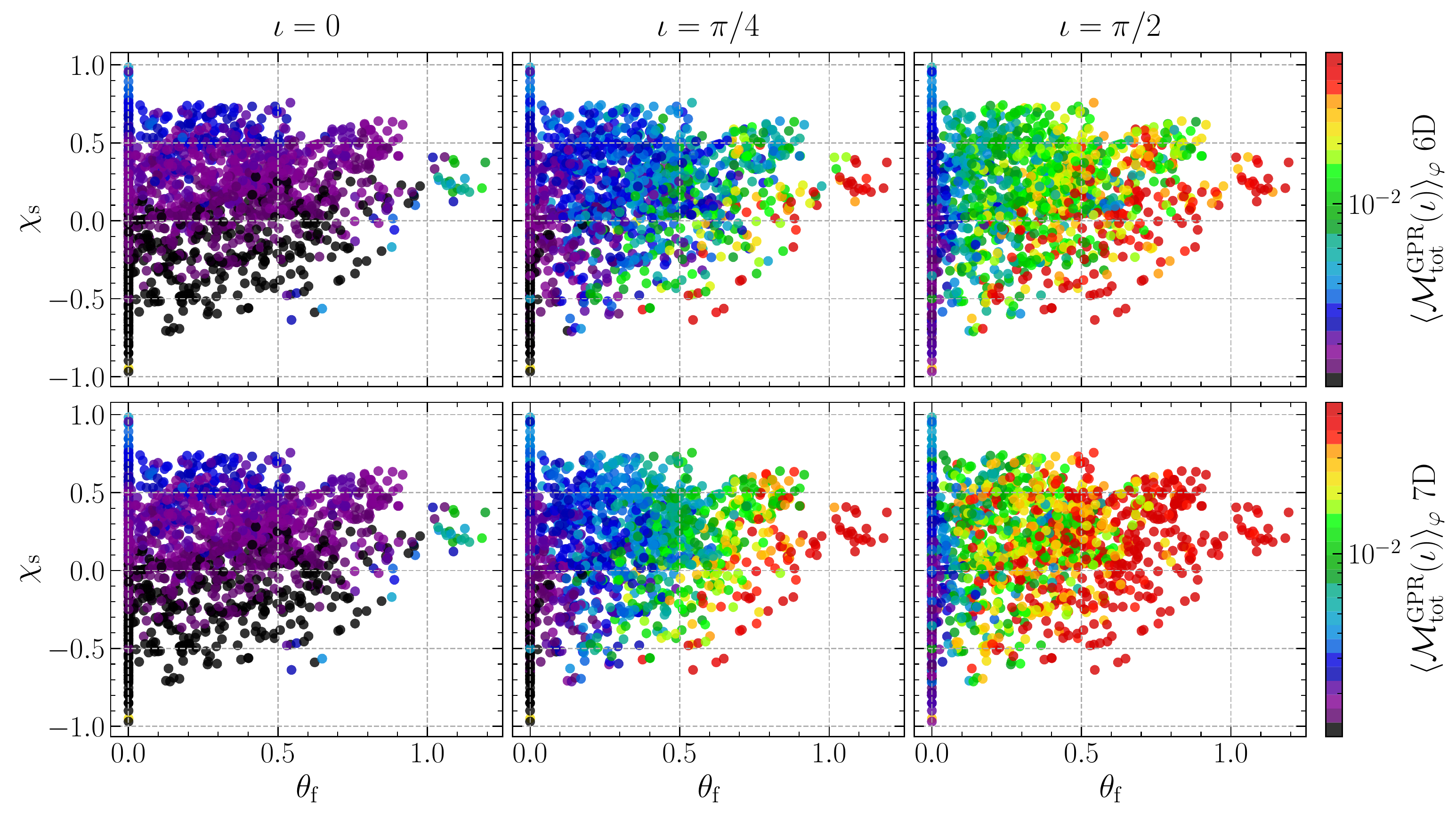}
    \caption{%
    Variations of the mismatch $\langle\mismGPRtot\rangle_\varphi$ for the 6D (top panels) and 7D (bottom panels) models as functions of the symmetric aligned-spin $\chis$ and the direction of the remnant spin $\thetaf$. Left, middle, and right panels refer to different inclination angles $\iota$.
}
    \label{fig:mismatch_scatterplot}
\end{figure*}

Results are presented in Table~\ref{tab:tot_mismatch} and Fig.~\ref{fig:violin_tot_waveform_eps_mism_4panel}, where we also include a comparison with error statistics minimized over $\varphi$. We find that $\langle\epsNRtot(\iota)\rangle_\varphi$ remains mostly between $10^{-3}$ and $10^{-1}$, consistent with the range of $\epsNR$ reported in Table~\ref{tab:LOO_err} and Fig.~\ref{fig:fits_3panel}.
We observe a reduction in performance across all listed metrics at higher inclination angles. This is expected: for $\iota = 0$ or $\iota = \pi$, only the $(2,\pm2)$ modes contribute, whereas at intermediate inclinations, additional modes become active, increasing waveform complexity.  Note also how the distributions of $\varphi$-averaged metrics differ more significantly from the $\varphi$-minimized ones: this indicates increased variability in the error as more modes contribute to the signal.
The distributions of the error metrics span the same orders of magnitude for both the 6D and 7D models, although the former consistently outperforms the latter.
Even though the distributions of $\epsGPRtot$ and $\mismGPRtot$ extend to higher values, the bulk of the distribution remains within the same range as $\epsNRtot$ and $\mismNRtot$. This indicates that both models can provide reliable predictions for data analysis.

\begin{figure}
    \centering
    \includegraphics[width=\columnwidth]{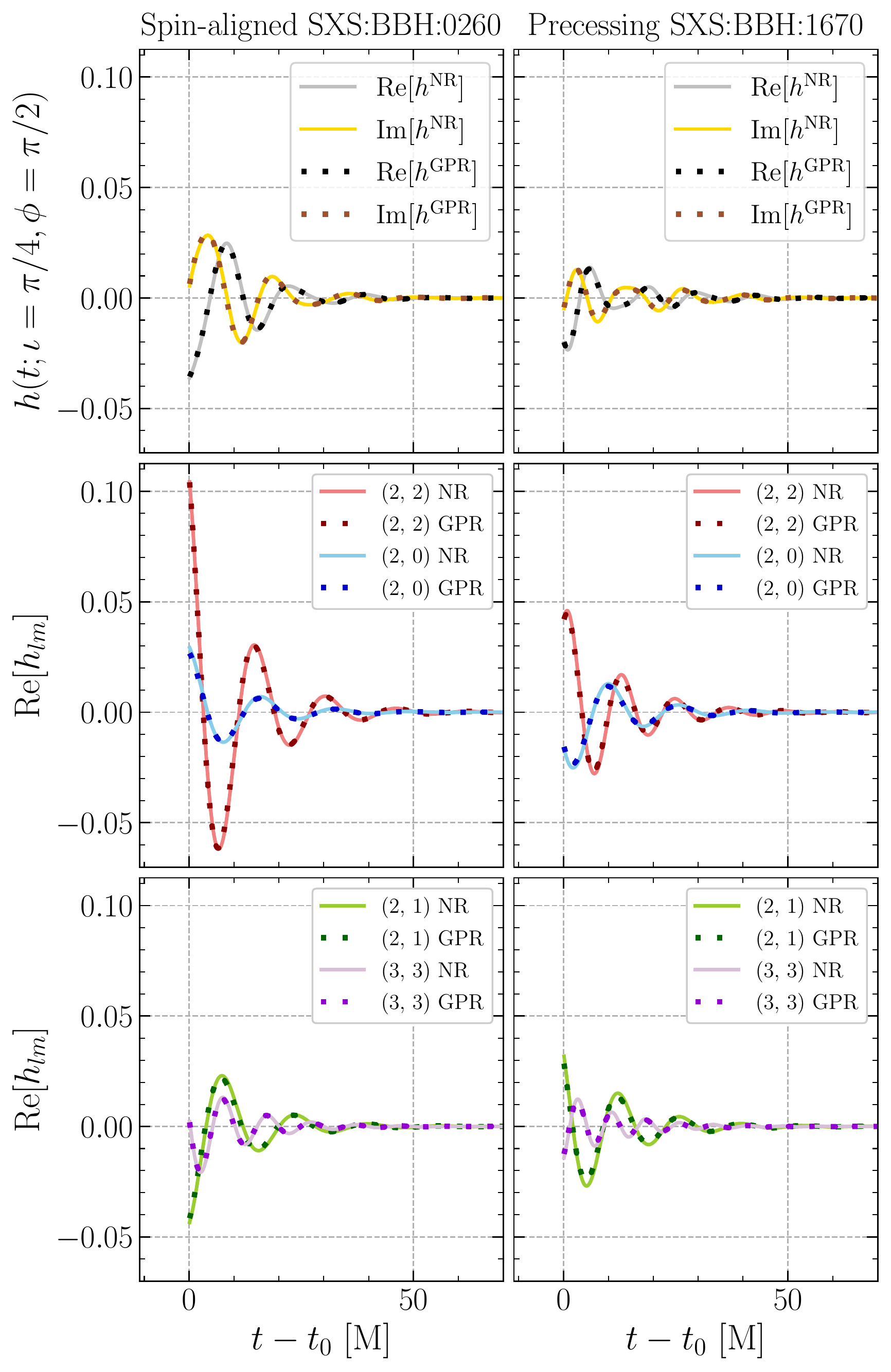}

    \caption{
Examples of ringdown waveforms reconstructed using our GPR fit. We consider two cases: a spin-aligned source (SXS:BBH:0260) in the left column, and a precessing source (SXS:BBH:1670) in the rigth one. The top row shows the strain obtained by combining the modes with Eq.~\eqref{eq:mode_combination} fixing $(\iota=\pi/4, \varphi=\pi/2)$, while the middle and bottom rows present the contribution of each individual mode. Solid curves indicate the waveforms extracted from SXS simulations, whereas dotted lines show our reconstruction using damped sinusoids. In this example, the mode amplitudes $A_{lm}$ are predicted by our 6D GPR model (leaving out the selected source from training), and the mode phases $\phi_{lm}$ are set to the values obtained from the waveform fitting procedure described in Sec.~\ref{sec:amp_extraction}.}
    \label{fig:waveform_example_combinedmodes}
\end{figure}

Figure~\ref{fig:mismatch_scatterplot} illustrates the correlations between $\langle\mismGPRtot\rangle_\varphi$ and the parameters $\thetaf$ and $\chis$ for both 6D and 7D models at inclination angles $\iota = 0, \pi/4, \pi/2$. Our analysis reveals that systems with higher values of $\langle\epsGPRtot\rangle_\varphi$ and $\langle\mismGPRtot\rangle_\varphi$ in both models tend to have larger values of $\thetaf$. Additionally, in the $\iota = 0$ case, we report that the mismatch increases as $\chis$ grows.

Figure~\ref{fig:waveform_example_combinedmodes} presents two example waveforms reconstructed from our fits. In particular, SXS:BBH:0260 is a spin-aligned system with $\thetaf=0$, $q=3.0$, $\chis=-0.85$, $\chia=-0.42$, $\phik=\pi/2$, $v_{\rm k}=0.0009$ and SXS:BBH:1670 is a precessing system with $\thetaf=0.77$, $q=3.2$, $\chis=0.39$, $\chia=0.27$, $\phik=0.28$, $v_{\rm k}=0.004$. To reconstruct the signals, we estimate the mode amplitude values using 6D fits that were trained using a dataset without these two waveforms. We observe that each mode amplitude is accurately predicted.  When combining the modes assuming veiwing angles $(\iota = \pi/4, \varphi = \pi/2)$, we find $\mismGPRtot = 0.00166, \epsGPRtot = 0.00333$ for the spin-aligned system, and $\mismGPRtot = 0.00168, \epsGPRtot = 0.00565$ for the precessing one. 

\section{Conclusions}
\label{sec:conclusions}

In this work, we investigated the phenomenology of BH ringdown from precessing, quasi-circular binaries and presented a first ever fit %
for their ringdown amplitudes. We focused on the $(2,\pm2)$, $(2,\pm1)$, $(3,\pm3)$, and $(2,0)$ modes, training our models on numerical relativity simulations from SXS catalog~\cite{Boyle:2019kee}.

We find that the ringdown amplitudes of precessing binaries exhibit a relatively smooth behavior in the 6D parameter space defined by $(\delta, \chis, \chia, \thetaf, \phik, v_{\rm k})$. Our key results on the phenomenological trends in the amplitudes of the ringdown modes are summarized in Table~\ref{tab:trend_summary}. Notably, the $(2,\pm1)$, $(3,\pm3)$, and $(2,0)$ modes can become comparable to the dominant modes for higher values of the misalignment angle $\thetaf$.

Our GPR models trained in this 6D subspace outperform those trained in the full 7D space defined by mass ratio and component spin magnitudes $(\delta, \chi_{1x,1y,1z},\chi_{2x,2y,2z})$. That said, the latter space is undoubtedly more familiar and closer to what is needed in current GW data-analysis pipelines. A possible avenue for future improvement is to explicitly model the transformation from the full 7D parameter space to the restricted 6D parameter space, and to assess whether this enhances the accuracy of the 7D model while preserving its ease of integration with current pipelines.
Alternatively, one could opt for data-driven dimensionality reduction schemes, which have already been successfully applied in other areas of GW physics, e.g. Refs. \cite{Saleem:2021nsb, Datta:2022izc,Engels:2014nua,Ferreira:2024gzh}. 
As a by-product, we also provide a GPR model for the quantity $\Delta\temop$, which predicts the reference time used to extract amplitude values from NR waveforms. All trained models are available in the \textsc{postmerger} package~\cite{Pacilio_postmerger_code}.

Both amplitude models produce mismatches $\lesssim 0.02$. Mismatches help evaluate the suitability of our results for gravitational-wave data analysis. Specifically, for a signal with a given signal-to-noise ratio (SNR), the impact of mismodeling between two waveform approximants with mismatch $\mathcal{M}$ becomes significant if~\cite{Lindblom:2008cm,Chatziioannou:2017tdw}
\begin{equation}
\mathcal{M} \gtrsim \frac{D}{2 \,{\rm SNR}^2}
\label{eq:mismatch}
\end{equation}
where $D$ is the number of parameters entering the waveform. When mismodeling errors surpass measurement uncertainty, addressing the systematic bias introduced by modeling limitations becomes essential, necessitating more precise models in data analysis. Equation (\ref{eq:mismatch}) provides a useful guideline for determining whether our fits achieve the required accuracy for a given data quality. With current detectors, ringdown signals have been measured at a SNRs $\lesssim 10$~\cite{Carullo:2019flw,Capano:2021etf,Ghosh:2021mrv,Siegel:2023lxl,LIGOScientific:2020tif,LIGOScientific:2021sio}. However, next-generation detectors are expected to observe a few events each year with SNRs around 100~\cite{Berti:2016lat,Bhagwat:2023jwv,Pacilio:2023mvk}. Furthermore, rough estimates using the Fisher matrix formalism~\cite{Bhagwat:2023jwv} suggest that a detector network consisting of Einstein Telescope and Cosmic Explorer could measure $A_{22}$ with sub-percent accuracy from approximately 8–10 events per year. Measurements of subdominant modes, however, would still retain a few percent uncertainty. Given these projections, while our current model may be sufficient for the present data quality, further refinements will be necessary to maintain accuracy as detector sensitivity improves and next-generation data becomes available.

This is our first attempt to produce ringdown amplitude fits for precessing binaries, and there is certainly room for improvement. We find that with our procedure, one key limiting factor is the size and parameter-space coverage of the NR training dataset. In the short term, we plan to improve our models by considering more NR catalogs, including the recently released third SXS catalog~\cite{Scheel:2025jct}, and those from the Rochester Institute of Technology group~\cite{Healy:2022wdn} and the MAYA catalog~\cite{Ferguson:2023vta}.
Exploring regression methods other than GPR that could make more efficient use of sparse datasets is also an important avenue for future work. Additional modes need to be incorporated, including spherical-spheroidal mixing modes, quadratic modes, and overtones. Finally, while in this work we have focused on predicting ringdown mode amplitudes, we aim to extend them to predict their phases in a follow-up work.

\begin{figure*}
 \centering
 \includegraphics[width=1.8\columnwidth]{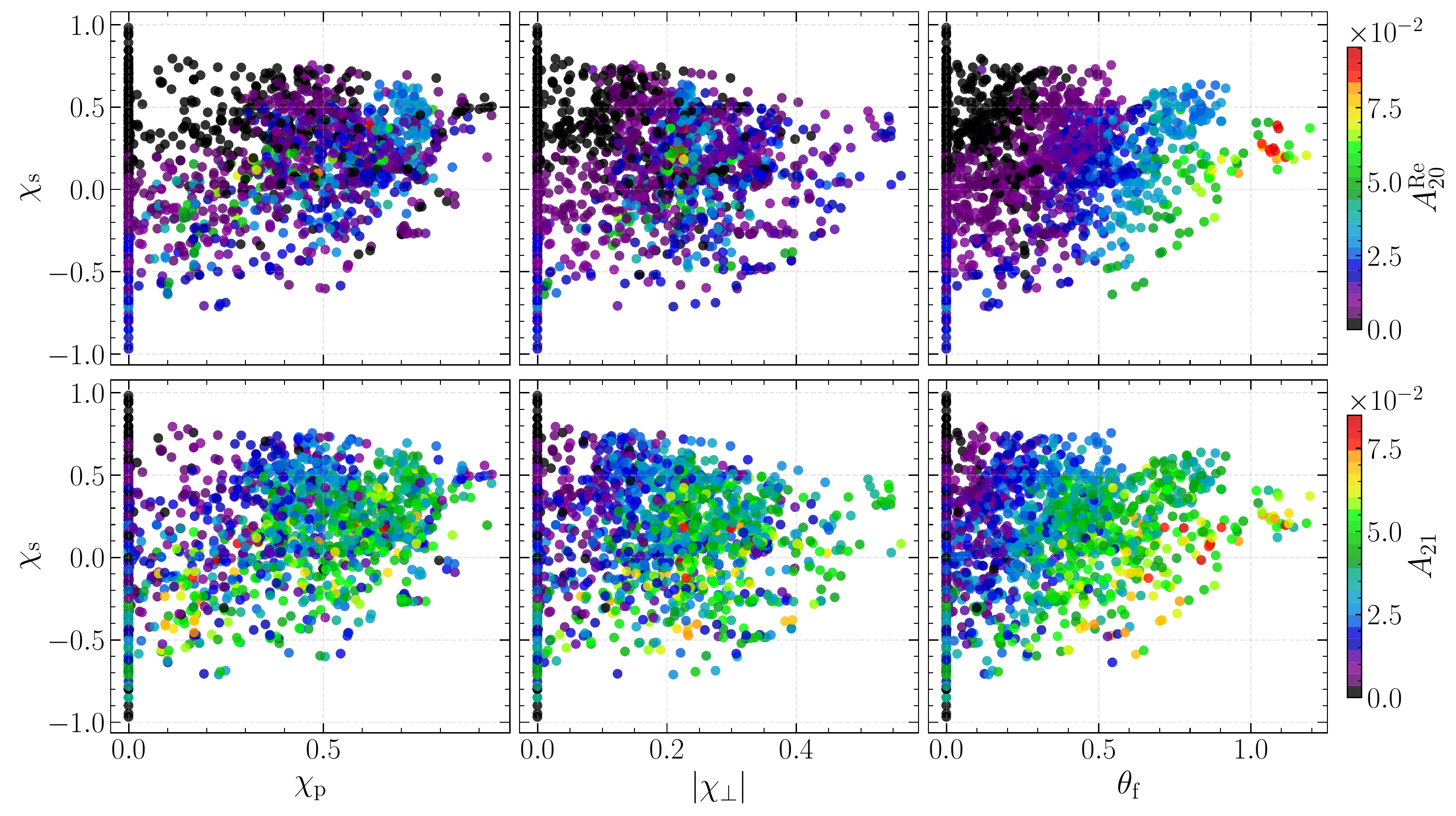}
 \caption{QNM amplitude values for the $(2,1)$ and Re$(2,0)$ modes are shown for both spin-aligned and precessing sources as a function of different spin-precession parameters. This comparison highlights how the final spin angle $\thetaf$ leads to smoother variations in the QNM amplitudes, in contrast to both the $\chi_{\rm p}$ parameter from Ref.~\cite{Gerosa:2020aiw} and the $|\boldsymbol{\chi}_\perp|$ parameter from Ref.~\cite{Thomas:2020uqj}.}
 \label{fig:chip_thetaf_comparison_scatterplots}
\end{figure*}

\section*{Acknowledgments}
We thank Mark Ho-Yeuk Cheung, Patricia Schmidt, Lucy Thomas, Mark Hannam, Xisco Jiménez Forteza, Sascha Husa, Gregorio Carullo, 
Lorena Magaña Zertuche, Leo Stein, and Federico Pozzoli for discussions. 
S.B. would like to acknowledge the UKRI Stephen Hawking Fellowship funded by the Engineering and Physical Sciences Research Council (EPSRC) with grant reference number EP/W005727 for support during this project.
C.P. and D.G. are supported by 
ERC Starting Grant No.~945155--GWmining, 
Cariplo Foundation Grant No.~2021-0555, 
MUR PRIN Grant No.~2022-Z9X4XS, 
MUR Grant ``Progetto Dipartimenti di Eccellenza 2023-2027'' (BiCoQ),
and the ICSC National Research Centre funded by NextGenerationEU. 
D.G. is supported by MSCA Fellowships No.~101064542--StochRewind and No.~101149270--ProtoBH.
Computational work was performed at CINECA with allocations 
through INFN and Bicocca,
and on the Bluebear %
cluster at the University of Birmingham. 

\section*{Data availability}
Trained models, supporting datasets, and examples are released as part of the \textsc{postmerger} open-source package for the python programming language, see Ref.~\cite{Pacilio_postmerger_code}.

\appendix

\section{Alternative parametrizations}
\label{app:choice_param_space}

We explored various combinations of physical parameters to identify smooth variations in the QNM amplitudes. In this appendix, we detail some of the attempts that led to the definition of our 6D parameter space.

The mass ratio $q$ is a key parameter in describing QNM phenomenology. However, rather than using $q$ directly, we opted for the variable $\delta = (q-1)/(q+1)$. This transformation provides a more uniform distribution for systems in the SXS catalog, improving both the visualization of amplitude variations and the performance of GPR training, as highlighted in Ref.~\cite{Pacilio:2024tdl}.

We also examine different physical quantities that encode information about the precession state of the systems. Specifically, we tested two alternative parametrizations proposed in the recent literature: the ``averaged'' definition of the $\chi_{\rm p}$ parameter introduced in Ref.~\cite{Gerosa:2020aiw} and the norm of the two-dimensional $\boldsymbol{\chi}_\perp$ parameter introduced in Ref.~\cite{Thomas:2020uqj}. Some results are presented in Fig.~\ref{fig:chip_thetaf_comparison_scatterplots}, which shows that neither of these parameters yields a smooth behavior for the ringdown amplitudes.

Motivated by Ref.~\cite{Zhu:2023fnf}, we instead opted for the misalignment angle $\thetaf$ between the remnant spin and the orbital angular momentum at ISCO, which provides a smoother behavior, as illustrated in Fig.~\ref{fig:chip_thetaf_comparison_scatterplots}.

\section{Definition and phenomenology of $\temop$}
\label{app:t_emop}

As described in Sec.~\ref{sec:amp_extraction}, in our fitting procedure, we choose $\temop + 20M$ as the reference time for the start of the ringdown phase. We define $\temop$ following Refs.\cite{Berti:2007fi,Baibhav:2017jhs}, but instead of using only one mode, we combine the effects of all $(2,m)$ modes. Specifically, for each of the $(2,m)$ modes,\footnote{Since the $(2,0)$ mode is not circularly polarized, we compute $E_{||,20}(t_0)$ separately for the real and imaginary parts.} we compute the parallel energy on a grid of integral start times $t_i$:
\begin{equation}
    E_{||,lm}(t_i) = \frac{1}{8\pi} \frac{\left| \int_{t_i} \dot{h}_{\mathrm{NR},lm} \dot{h}_{\mathrm{QNM},lm}^* dt \right|^2}{\int_{t_i} \dot{h}_{\mathrm{QNM},lm} \dot{h}_{\mathrm{QNM},lm}^* dt}\,.
    \end{equation}
   The total parallel energy is given by the quadrature sum
       \begin{equation}
    E_{||,l=2}(t_i) = \sqrt{\sum_m E_{||,2m}(t_i)}\,,
   \end{equation}
and can be interpreted as the energy content in the ringdown portion of the waveform. Now, $\temop$ is the time at which this energy is maximized:
\begin{align}
   E_{\rm EMOP} &= E_{||,l=2}(t_{\mathrm{EMOP}}) = \max\limits_{t_i} E_{||,l=2}(t_i)\,, \\
    t_{\rm EMOP} &= \arg \max_{t_{i}} E_{||,l=2}(t_i)\,.
\end{align}

To study how $\temop$ changes as a function of the binary parameters, we introduce a second reference time, $t_{\rm peak\_norm} = \max\limits_{t} \sqrt{\sum_{l,m}\left| h_{l,m} \right|^2}$, which corresponds to the peak of the $L^2$ norm of the waveforms, considering all modes present in the simulations. We then define $\Delta t_{\rm EMOP} = t_{\rm EMOP} - t_{\rm peak\_norm}$.

\begin{figure}
     \centering
     \includegraphics[width=\columnwidth]{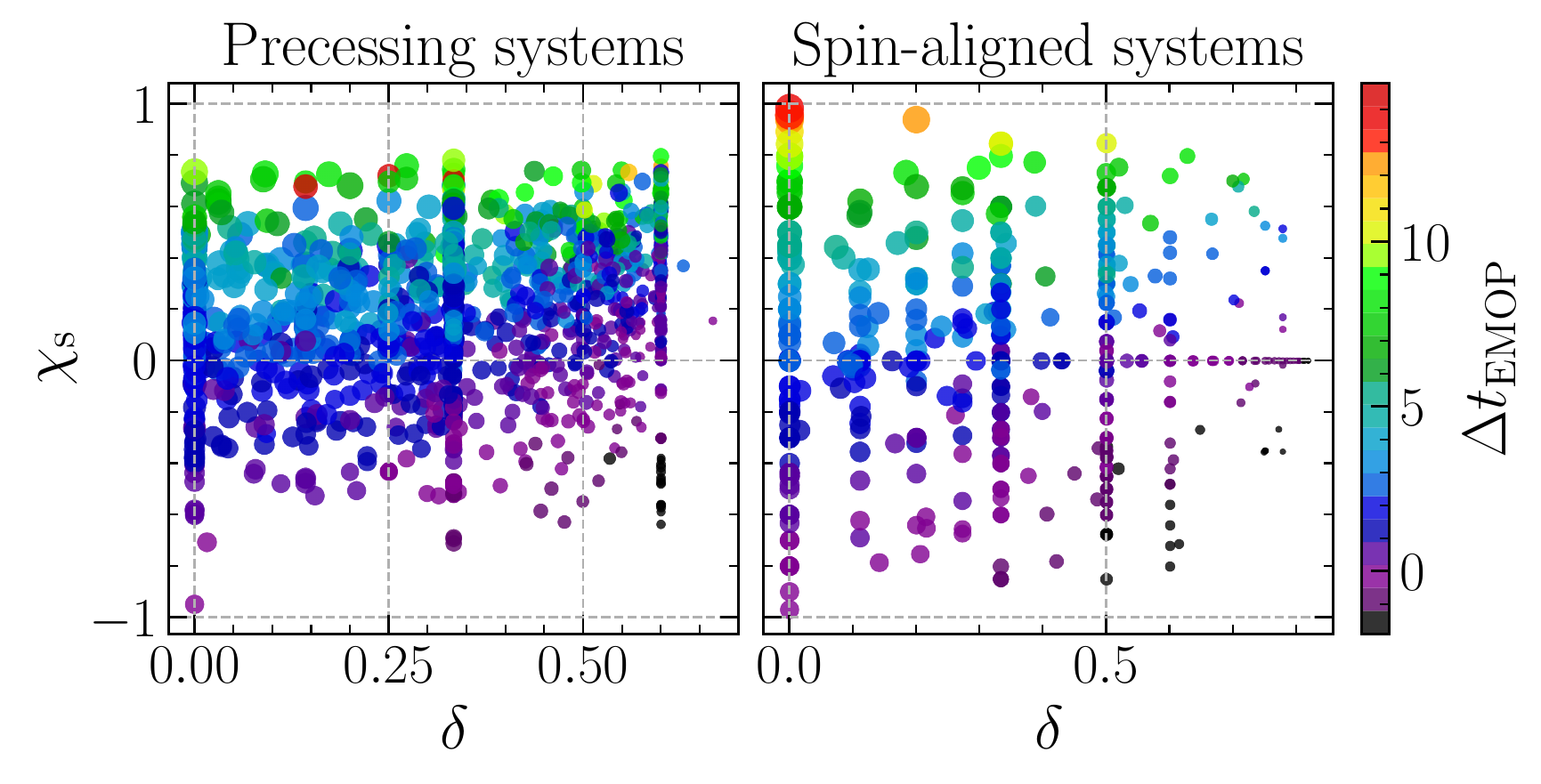}
     \caption{Values of $\Delta t_{\rm EMOP}$ (represented by marker color) and $E_{\rm EMOP}$ (represented by marker size) as a function of $\chis$ and $\delta$ for precessing (left) and spin-aligned (right) systems in our dataset. Both $\Delta t_{\rm EMOP}$ and $E_{\rm EMOP}$ decrease for systems with higher mass ratios and negative $\chis$.}
     \label{fig:t_emop_scatter}
 \end{figure}

For our dataset, we have $-2.3M < \Delta t_{\rm EMOP} < 14.8M$ and $1.2 \times 10^{-3} M < E_{\rm EMOP} < 2.97 \times 10^{-2} M$. We observe that $\chis$ and $\delta$ are the most dominant parameters along which $\Delta t_{\rm EMOP}$ and $E_{\rm EMOP}$ vary smoothly. This is shown in Fig.~\ref{fig:t_emop_scatter}, for both precessing and spin-aligned systems. Both $\Delta t_{\rm EMOP}$ and $E_{\rm EMOP}$ decrease for systems with larger mass ratios and negative $\chis$. Along with the GPR models for mode amplitudes, we also trained 6D and 7D models for $\Delta t_{\rm EMOP}$, providing handy tools to estimate $\temop$ from binary parameters. Performance metrics computed using the LOO framework for these models are reported in Table \ref{tab:dt_emop}.

\begin{table}
\centering
\renewcommand{\arraystretch}{1.3} %
\begin{tabular}{c|c||c|c}
Subsample & Error & 6D & 7D \\
\hline \hline
\multirow{2}{*}{All systems}  
    & Abs. & $0.20^{+1.58}_{-0.19}$ & $0.36^{+2.56}_{-0.35}$ \\
    & Rel. & $0.12^{+1.13}_{-0.12}$ & $0.18^{+4.48}_{-0.18}$ \\
\hline
\multirow{2}{*}{$|\Delta t_{\rm EMOP}| > 2M$}  
    & Abs. & $0.34^{+2.06}_{-0.32}$ & $0.35^{+3.57}_{-0.34}$ \\
    & Rel. & $0.087^{+0.424}_{-0.083}$ & $0.098^{+0.602}_{-0.096}$ \\
\hline
\multirow{2}{*}{Precessing}  
    & Abs. & $0.34^{+1.76}_{-0.32}$ & $0.76^{+2.46}_{-0.74}$ \\
    & Rel. & $0.19^{+1.33}_{-0.18}$ & $0.34^{+5.38}_{-0.33}$ \\
\hline
\multirow{2}{*}{Spin-aligned}  
    & Abs. & $0.051^{+0.206}_{-0.047}$ & $0.042^{+0.084}_{-0.038}$ \\
    & Rel. & $0.030^{+0.277}_{-0.027}$ & $0.018^{+0.299}_{-0.016}$ \\
\end{tabular}
\caption{Performance of our GPR fit for $\Delta t_{\rm EMOP}$ considering different subset of systems. Time is expressed in units of total mass $M$.}
\label{tab:dt_emop}
\end{table}

\section{Retrograde modes}
\label{app:retrograde_modes}

In our fits, we have neglected the effects of retrograde modes. In this appendix, we quantify to what extent this assumption holds.

\begin{figure}
    \centering
    \includegraphics[width=\columnwidth]{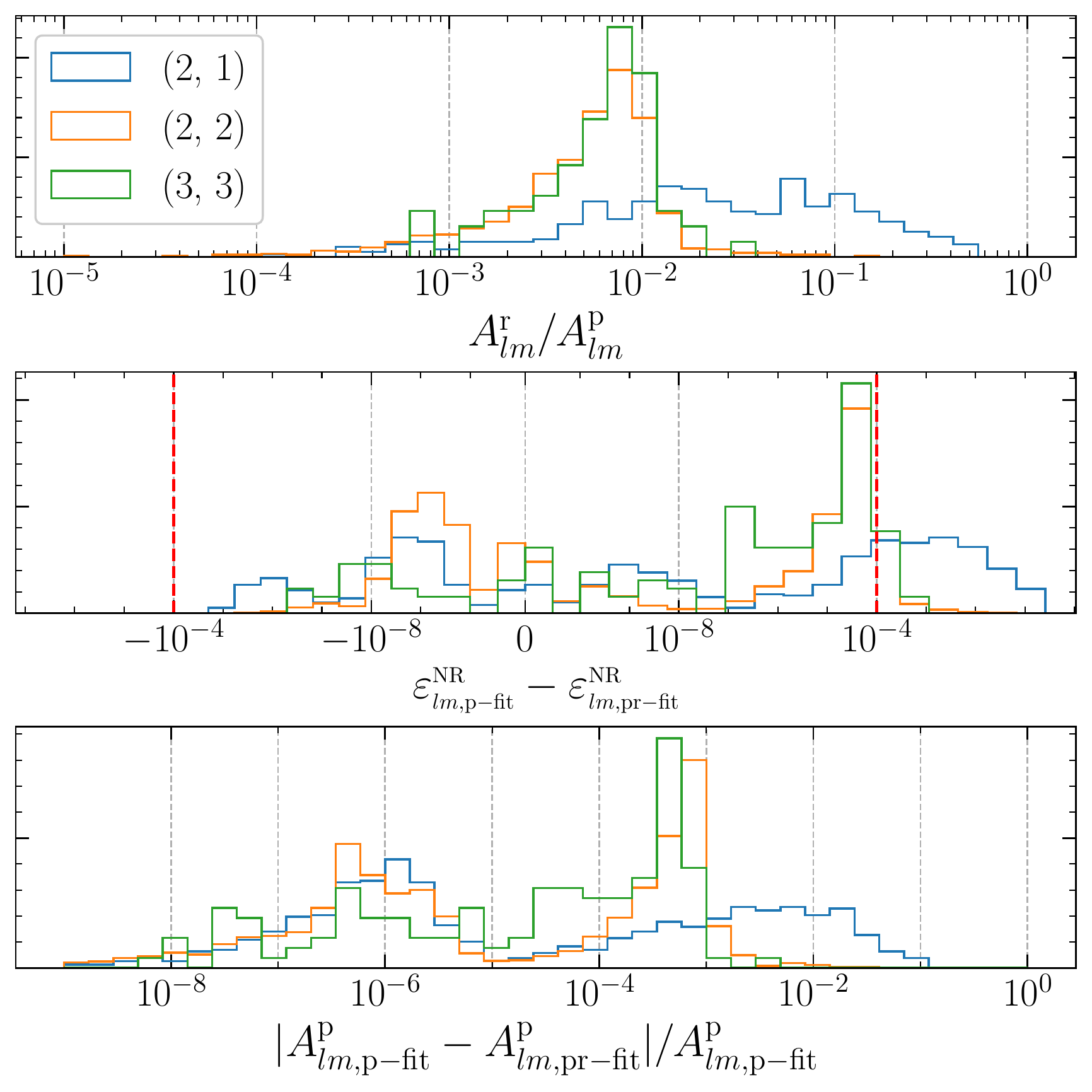}
    \caption{Properties of the retrograde-mode ringdown amplitude for $A_{lm}>0.03$. Results for the $(2,1)$, $(2,2)$, and $(3,3)$ modes are reported in blue, orange, and green, respectively. The top panels show the ratio between prograde and retrograde mode amplitudes. The middle panel shows the difference in the fitting error between prograde-only fitting and prograde+retrograde waveform fitting; red dashed lines mark the threshold $\epsNR = 10^{-4}$. The bottom panel shows the relative difference in prograde amplitude values between prograde-only and prograde+retrograde waveform fitting.}
    \label{fig:retrog_histograms}
\end{figure}

We define retrograde modes following the same convention as in Refs.~\cite{Cheung:2023vki, Pacilio:2024tdl}, and we repeat the waveform fitting process described in Sec.~\ref{sec:amp_extraction} with a different ansatz:
\begin{align}
\label{eq:waveform_proret}
    h_{lm}(t) = &A_{lm}^{\mathrm{p}} \, e^{-(t-t_0)/\tau_{lm}^{\mathrm{p}}} e^{-i\omega_{lm}^{\mathrm{p}}(t-t_0) + \phi_{lm}^{\mathrm{p}}} \nonumber \\
    &+ A_{lm}^{\mathrm{r}} \, e^{-(t-t_0)/\tau_{lm}^{\mathrm{r}}} e^{-i\omega_{lm}^{\mathrm{r}}(t-t_0) + \phi_{lm}^{\mathrm{r}}}
\end{align}
Results are presented in Fig.~\ref{fig:retrog_histograms}, where we restrict to systems with $A_{lm}>0.03$.

In the top panel of Fig.~\ref{fig:retrog_histograms}, we report the distributions of the ratio $A_{lm}^{\mathrm{r}}/A_{lm}^{\mathrm{p}}$ for the $(2,2),(2,1),(3,3)$ modes (results are similar for the $m<0$ modes). For $(2,\pm2)$ and $(3,\pm3)$, the bulk of the distribution remains below $10^{-2}$, with very small tails extending up to $10^{-1}$: for these modes, retrograde amplitudes remain significantly subdominant. The relative retrograde excitation can be slightly more pronounced for the $(2,\pm1)$ modes, reaching a few percent of the prograde amplitude in around $25\%$ of the simulations. In a few cases, the ratio can be as high as $50\%$.

In the middle panel of Fig.~\ref{fig:retrog_histograms}, we examine the difference in fitting error when switching from a prograde-only fit ($\varepsilon^{\rm NR}_{lm, \rm{p-fit}}$) to a prograde+retrograde fit ($\varepsilon^{\rm NR}_{lm, \rm{pr-fit}}$). We set a significance threshold of $\pm 10^{-4}$, which corresponds to the lower limit for $\varepsilon^{\rm NR}$ as discussed in Sec.~\ref{sec:amp_extraction}. When moving to the prograde+retrograde fit, we observe no significant decrease in the fitting error. For the $(2,\pm2)$ and $(3,\pm3)$ modes, only about 5\% of the simulations show a difference $\Delta \varepsilon^{\rm NR}_{\scriptscriptstyle lm} > 10^{-4}$, meaning that, for these modes, the improvement from adding retrograde fitting is negligible. However, for the $(2,\pm1)$ modes, around 26\% of the simulations show $\Delta \varepsilon^{\rm NR}_{\scriptscriptstyle lm} > 10^{-4}$ and the improvement can reach up to $\Delta \varepsilon^{\rm NR}_{\scriptscriptstyle lm} \simeq 10^{-1}$.

In the bottom panel of Fig.~\ref{fig:retrog_histograms}, we investigate the relative variation in the extracted amplitude of the prograde mode when transitioning from the prograde-only fitting ($A^{\rm p}_{lm, \rm{p-fit}}$) to the prograde+retrograde fitting ($A^{\rm p}_{lm, \rm{pr-fit}}$). This variation is negligible for $(2,\pm2)$ and $(3,\pm3)$ modes, with the relative change always below 1\%. For the $(2,\pm1)$ modes, the variation in prograde amplitude can range from 1\% to 10\%.

\begin{figure}[t!]
    \centering
    \includegraphics[width=\columnwidth]{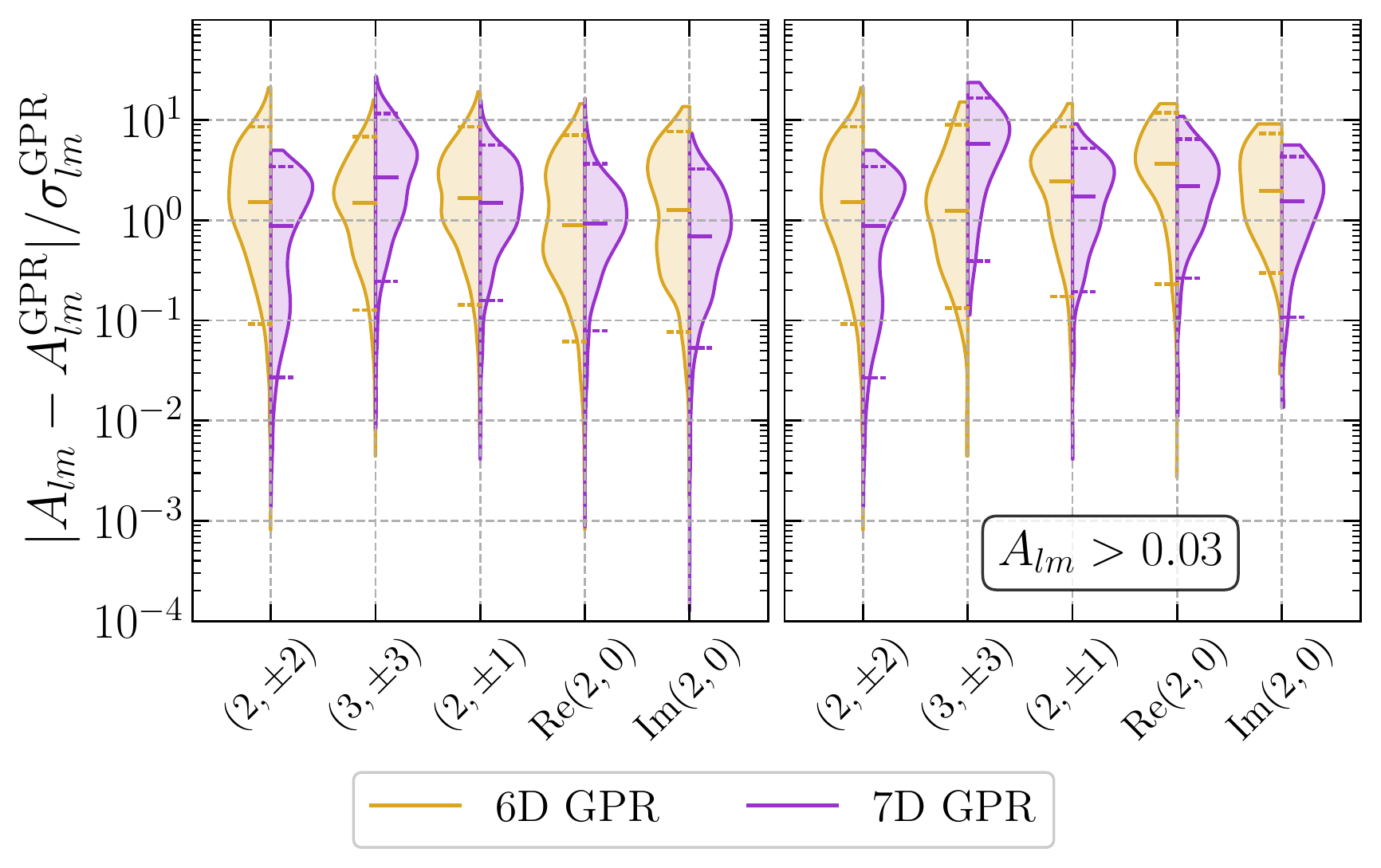}
    \caption{Ratio of absolute prediction errors via LOO and GPR uncertainties $\sigma^{\rm GPR}_{lm}$ for each mode. We report results for both our 6D (blue) and 7D (orange) parameter spaces. The left panel includes all training sources, while in the right panel we restrict to cases with $A_{lm} > 0.3$.}
    \label{fig:violin_norm_err_2panel}
\end{figure}

\section{GPR uncertainties}
\label{app:sigma_gpr}

Given a target point in the parameter space, a GPR model outputs a Gaussian distribution characterized by a mean value $y^{\rm GPR}$ and a standard deviation $\sigma^{\rm GPR}$. While $\sigma^{\rm GPR}$ provides an estimate of the uncertainty associated with $y^{\rm GPR}$, properly interpreting this estimate requires understanding the distinct sources of uncertainty.

Uncertainty in our models can be categorized into two types~\cite{hullermeier2021aleatoric,valdenegro2022deeper}: aleatoric uncertainty, which arises from intrinsic variability in the target quantity (amplitude values, in our case), and epistemic uncertainty, which is linked to the distribution of data points in the parameter space. Gaussian processes inherently model epistemic uncertainty through their kernel functions, which depend on the geometrical distance between data points. As a result, epistemic uncertainty is low near data points and high in sparse regions.

The simplest model for aleatoric uncertainty is additive white noise with a constant magnitude throughout the domain. In our implementation, we attempt to account for this by including a white noise kernel in the covariance function. However, our homoscedastic model fails to capture regions with varying amplitude variance, such as those distinguishing precessing and spin-aligned systems. The constant white noise kernel is ineffective in modeling noise in our dataset and is effectively "turned off" during training [i.e., the marginalized likelihood reaches its minimum when the $N_0$ parameter in Eq.\eqref{eq:kernel} is at its lower boundary].\footnote{The only exception is the 7D GPR model for the $(2,2)$ mode, where the white noise kernel is activated. Nevertheless, the resulting error estimate remains inaccurate, as shown in Fig.~\ref{fig:violin_norm_err_2panel}.}

As a result, $\sigma^{\rm GPR}$ reflects only epistemic uncertainty and does not provide a reliable estimate of the total prediction error. To quantify this limitation, in Fig.~\ref{fig:violin_norm_err_2panel}, we analyze the ratio between the LOO absolute prediction error (cf. Section \ref{sec:GPR_setup}) and $\sigma^{\rm GPR}$. We find that this ratio spans two orders of magnitude around 1, indicating that the nominal GPR error can both significantly under- and overestimate the more reliable LOO error.

More accurate treatment of uncertainty requires heteroscedastic GPRs, which can model aleatoric uncertainty varying across the parameter space~\cite{goldberg1997regression,kersting2007most}. We leave this for future work.

\bibliography{precessing_ringdown}
\end{document}